\RequirePackage{etoolbox}
\csdef{input@path}{{style/}{graphics/}}
\documentclass[MSNbibl,nameyear,seceqn,dvips]{arxstspdf}
\usepackage{algorithm}
\usepackage{graphicx}
\usepackage{url,breakurl}
\usepackage{flushend}
\usepackage{stfloats}

\volume{29}
\issue{4}
\pubyear{2014}
\firstpage{485}
\lastpage{511}
\doi{10.1214/14-STS500} % kopijuoti is laisko
\docsubty{FLA}

\makeatletter

\newcommand{\rrvert}{\vert}
\newcommand{\llvert}{\vert}
\newcommand{\R}{\mathbb{R}}
\newcommand{\eps}{\varepsilon}
\newcommand{\A}{\mathcal{A}}
\newcommand{\X}{\mathcal{X}}
\newcommand{\q}{\varrho}
\newcommand{\qmax}{\varrho_{\mathrm{max}}}
\newcommand{\st}{\dvtx}
\newcommand{\hr}{\hat{r}}
\newcommand{\hl}{\hat{l}}
\newcommand{\tx}{\tilde{x}}
\newcommand{\vs}{versus}%{\textit{vs.}}
\newcommand{\hmu}{\hat{\mu}}
\newcommand{\hpi}{{\hat{\pi}}}
\newcommand{\piPE}{\pi_{\mathrm{PV}}}
\newcommand{\event}{\mathcal{E}}
\newcommand{\ind}{I}
\renewcommand{\Pr}{\mathbf{P}}
\newcommand{\accept}{\operatorname{accept}}
\newcommand{\reach}{\operatorname{reach}}

\newcommand{\rcv}{\texttt{rcv1}}
\newcommand{\rmse}{\textsf{rmse}}
\newcommand{\bias}{\textsf{bias}}
\newcommand{\std}{\textsf{stdev}}

\newcommand{\EE}{\mathop{\E}}
\newcommand{\VV}{\mathop{\Var}}
\newcommand{\PP}{\Pr}

\newcommand{\E}{\mathbf{E}}
\newcommand{\Var}{\mathbf{V}}

\newtheorem{theorem}{Theorem}[section]
\newtheorem{lemma}[theorem]{Lemma}
\newtheorem{lemmas}{Lemma}
\newproclaim{assumption}[theorem]{Assumption}

\newcommand{\hV}{\hat{V}}
\newcommand{\DR}{{\mathrm{DR}}}
\newcommand{\DRns}{{\mathrm{DR\mbox{-}ns}}}
\newcommand{\WC}{{\mathrm{DR\mbox{-}ns\mbox{-}wc}}}
\newcommand{\DM}{{\mathrm{DM}}}
\newcommand{\IPS}{{\mathrm{IPS}}}
\newcommand{\argmax}{\operatorname{argmax}}
\renewcommand{\rho}{q}
\newcommand{\coloneqq}{:=}
\newcommand{\init}{\mathrm{init}}
\newcommand{\valid}{\mathrm{valid}}
\newcommand{\eval}{\mathrm{eval}}
\renewcommand{\emptyset}{\varnothing}

\makeatother

\begin{document}
\begin{frontmatter}

\title{Doubly Robust Policy Evaluation and Optimization\thanksref{T1}}
% kai straipsnis turi susijusiu diskusiju ir rejoinder'iu
%rejoinder at \relateddoi{r}{10.1214/00-STSXXXX}.}
\runtitle{Doubly Robust Policy Evaluation and Optimization}

\begin{aug}
\author[A]{\fnms{Miroslav} \snm{Dud\'ik}\ead[label=e1]{mdudik@microsoft.com}},
\author[B]{\fnms{Dumitru} \snm{Erhan}\ead[label=e2]{dumitru@google.com}},
\author[A]{\fnms{John} \snm{Langford}\ead[label=e3]{jcl@microsoft.com}}
\and
\author[C]{\fnms{Lihong} \snm{Li}\corref{}\ead[label=e4]{lihongli@microsoft.com}}

\runauthor{Dud\'ik, Erhan, Langford and Li}

\affiliation{Microsoft Research and Google Inc.}

\address[A]{Miroslav Dud\'ik is Senior Researcher and John Langford is
Principal Researcher,
Microsoft Research, New York, New York,
USA (e-mail: \printead*{e1}; \printead*{e3}).}
\address[B]{Dumitru Erhan is Senior Software Engineer, Google Inc.,
Mountain View,
California, USA \printead{e2}.}
\address[C]{Lihong Li is Researcher, Microsoft Research, Redmond,
Washington, USA 
\printead{e4}.}
\thankstext[1]{T1}{Parts of this paper were presented at the 28th
International Conference on
Machine Learning (Dud{\'i}k, Langford and Li, \citeyear{DLL11}), and the 28th
Conference on Uncertainty
in Artificial Intelligence (Dud{\'i}k et~al., \citeyear{DELL12}).}
\end{aug}

% ABSTRACT
%
\begin{abstract}
We study sequential decision making in environments where rewards are only
partially observed, but can be modeled as a function of observed contexts
and the chosen action by the decision maker.
This setting, known as contextual bandits,
encompasses a wide variety of applications such as health care,
content recommendation
and Internet advertising. A central task is evaluation of a new
policy given historic data consisting of contexts, actions and
received rewards. The key challenge is that the past data typically
does not faithfully represent proportions of actions taken by a new
policy. Previous approaches rely either on models of rewards
or models of the past policy. The former are plagued by a large bias
whereas the latter have a large variance.

In this work, we leverage the strengths and
overcome the weaknesses of the two approaches by applying
the \emph{doubly robust} estimation technique to the problems of
policy evaluation and optimization. We prove that this
approach yields accurate
value estimates when we have \emph{either}
a good (but not necessarily consistent) model of rewards
\emph{or} a good (but not necessarily consistent) model of past
policy. Extensive empirical
comparison demonstrates that the doubly robust estimation uniformly
improves over existing
techniques, achieving both lower variance in value estimation and better
policies. As such, we
expect the doubly robust approach to become common practice in
policy evaluation and optimization.
\end{abstract}

% KEYWORDS
\end{frontmatter}

%s1 #&#
\section{Introduction}
\label{sec:introduction}

Contextual bandits (\citeauthor{EXP4}, \citeyear{EXP4}; 
Langford and Zhang, \citeyear{Epoch-Greedy}), sometimes known as
associative reinforcement learning (\citeauthor{Barto85Pattern},
\citeyear{Barto85Pattern}), are a
natural generalization of the classic multiarmed bandits introduced by
\citet{Robbins52Some}. In a contextual bandit problem, the decision
maker observes contextual information, based on which an action is
chosen out of a set of candidates; in return, a numerical ``reward''
signal is observed for the \emph{chosen} action, but not for others.
The process repeats for multiple steps, and the goal of the decision
maker is to maximize the total rewards in this process. Usually,
contexts observed by the decision maker provide useful information to
infer the expected reward of each action, thus allowing greater rewards
to be accumulated, compared to standard multi-armed bandits, which take
no account of the context.

Many problems in practice can be modeled by contextual bandits.
%This paper considers decision making problems where, at each step, we
%observe certain contextual features, choose one out of a set of
%candidate actions, and receive a numerical ``reward'' signal only for
%the chosen action. The goal is to maximize the sum of rewards over
%time.
For example, in one type of Internet advertising, the decision maker
(such as a website) dynamically selects which ad to display to a user
who visits the page, and receives a payment from the advertiser if the
user clicks on the ad (e.g., \citeauthor{Chapelle12Empirical},
\citeyear{Chapelle12Empirical}). In this
case, the context can be the user's geographical information, the
action is the displayed ad and the reward is the payment. Importantly,
we find only whether a user clicked on the
presented ad, but receive no information about the ads that were not presented.

Another example is content recommendation on Web portals (\citeauthor
{Agarwal13Content}, \citeyear{Agarwal13Content}). Here, the decision
maker (the web portal) selects,
for each user visit, what content (e.g., news, images, videos and
music) to display on the page. A natural objective is to
``personalize'' the recommendations, so that the number of clicks is
maximized (\citeauthor{Li10Contextual}, \citeyear{Li10Contextual}).
In this case, the context is the
user's interests in different topics, either self-reported by the user
or inferred from the user browsing history; the action is the
recommended item; the reward can be defined as $1$ if the user clicks
on an item, and $0$ otherwise.

Similarly, in health care, we only find out the clinical outcome (the
reward) of a patient who received a treatment (action), but not the
outcomes for alternative treatments. In general, the treatment strategy
may depend on the context of the patient such as her health level and
treatment history. Therefore, contextual bandits can also be a natural
model to describe personalized treatments.
%More discussions are found in \sec{related-dtr}.
%Both of these problems are instances of \emph{contextual bandits}
%information about the user or patient.

The behavior of a decision maker in contextual bandits can be described
as a \emph{policy}, to be defined precisely in the next sections.
Roughly speaking, a policy is a function that maps the decision maker's
past observations and the contextual information to a distribution over
the actions. This paper considers the
\emph{offline} version of contextual bandits: we assume access to
historical data, but
no ability to gather new data (\citeauthor{Langford08Exploration},
\citeyear{Langford08Exploration}; \citeauthor{ESII}, \citeyear{ESII}).
There are two related tasks that arise in this setting: \emph{policy
evaluation}
and \emph{policy optimization}.
%
% MD omitting: (or \emph{learning}).
%
The goal of policy evaluation
is to estimate the expected total reward of a \emph{given} policy. The
goal of policy optimization
is to obtain
%
% MD omitting:, or \emph{learn},
%
a policy that (approximately) maximizes expected total rewards. The
focus of this
paper is on policy evaluation, but as we will see in the experiments,
the ideas can
also be applied to policy optimization.
The offline version of contextual bandits is important in practice. For
instance, it allows a website to estimate, from historical log data,
how much gain in revenue can be achieved by changing the ad-selection
policy to a new one (\citeauthor{Bottou13Counterfactual}, \citeyear
{Bottou13Counterfactual}). Therefore, the
website does not have to experiment on \emph{real} users to test a new
policy, which can be very expensive and time-consuming. Finally, we
note that this problem is a special case of \emph{off-policy}
reinforcement learning (\citeauthor{Precup00Eligibility}, \citeyear
{Precup00Eligibility}).

Two kinds of approaches address offline policy evaluation.
The first, called the \emph{direct
method} (DM), estimates the reward function from given data and uses
this estimate in place of actual reward to evaluate the policy value
on a set of contexts. The second kind, called \emph{inverse propensity
score} (IPS) (\citeauthor{HorvitzTh52}, \citeyear{HorvitzTh52}), uses
importance weighting to correct
for the incorrect
proportions of actions in the historic data. The first approach requires
an accurate model of rewards, whereas the second approach requires an
accurate model of the past policy. In general, it might be difficult
to accurately model rewards, so the first assumption can be too
restrictive. On the other hand, in many applications, such as
advertising, Web search and content recommendation, the decision maker has
substantial, and possibly perfect, knowledge of the past policy, so the
second approach
can be applied. However, it often
suffers from large variance, especially when the past policy differs
significantly from the policy being evaluated.

In this paper, we propose to use the technique of \emph{doubly robust} (DR)
estimation to overcome problems with the two existing
approaches.
Doubly robust (or doubly protected)
estimation (Cassel, S{\"a}rndal and Wretman, \citeyear{CasselSaWr76};
Robins, Rotnitzky and Zhao, \citeyear{Robins94Estimation};
\citeauthor{RR95}, \citeyear{RR95}; \citeauthor{LD04}, \citeyear{LD04};
\citeauthor{KangSc07}, \citeyear{KangSc07})
is a statistical approach for
estimation from incomplete data with an important property: if \emph
{either one}
of the two estimators (i.e., DM or IPS) is correct, then the estimation
is unbiased.
This method thus increases the chances of drawing reliable inference.

%For example, when conducting a survey, seemingly ancillary questions
%such as age, sex, and family income may be asked. Since not everyone
%contacted responds to the survey, these values along with census
%statistics may be used to form an estimator of the probability of a
%response conditioned on age, sex, and family income. Using importance
%weighting inverse to these estimated probabilities, one estimator of
%overall opinions can be formed. An alternative estimator can be
%formed by directly regressing to predict the survey outcome given any
%available sources of information. Doubly robust estimation unifies
%these two techniques, so that unbiasedness is guaranteed if
%estimate is accurate \emph{or} the regressed predictor is accurate.

We apply the doubly robust technique to policy evaluation
and optimization in a contextual bandit setting.
The most straightforward policies to consider are \emph{stationary}
policies, whose actions depend on the current, observed context
alone. \emph{Nonstationary} policies, on the other hand, map the current
context and a history of past rounds to an action. They are of critical
interest because \emph{online learning} algorithms (also known as
adaptive allocation rules), by definition, produce nonstationary policies.
%the explore/exploit tradeoff inherent in
%this setting are, by definition, nonstationary policies.
We address both
stationary and nonstationary policies in this paper.

In Section~\ref{sec:related}, we describe previous work and connect
our setting to the related area of dynamic treatment regimes.

In Section~\ref{sec:stationary}, we study stationary policy
evaluation, analyzing the bias and variance of our core technique.
Unlike previous theoretical analyses, we do not assume that either the
reward model or
the past policy model are correct. Instead, we show how the deviations
of the two models from the truth impact bias and variance of the
doubly robust estimator. To our knowledge, this style of analysis is
novel and
may provide insights into doubly robust estimation beyond the specific
setting studied here. In Section~\ref{sec:experiment1}, we apply this
method to
both policy evaluation and optimization, finding that this approach can
substantially sharpen existing techniques.

In Section~\ref{sec:nonstationary}, we consider nonstationary policy
evaluation. The main approach
here is to use the historic data to obtain a sample of the run of an evaluated
nonstationary policy via rejection sampling (\citeauthor{LCLW11},
\citeyear{LCLW11}). We combine
the doubly robust technique with an improved form of rejection sampling
that makes
better use of data at the cost of small, controllable bias. Experiments
in Section~\ref{sec:experiment2} suggest the combination is able to
extract more information
from data than existing approaches.

%s2 #&#
\section{Prior Work} \label{sec:related}

%s2.1 #&#
\subsection{Doubly Robust Estimation} \label{sec:related-dr}

Doubly robust estimation is widely used in statistical inference
(see, e.g., \citeauthor{KangSc07}, \citeyear{KangSc07}, and the
references therein).
% \citep{Robins94Estimation} is a statistical
%technique addressing value estimation with nonrandomized
%%nonrepresentative
%sampling.
%For example,
More recently, it has been used in Internet advertising to
estimate the effects of new features for online
advertisers (Lambert and Pregibon, \citeyear{DRAds}; \citeauthor
{DRAds2}, \citeyear{DRAds2}).
%Much of previous work focuses on parameter estimation rather than
%policy evaluation and optimization, as addressed here. Furthermore,
Most of previous analysis of doubly robust estimation
is focused on asymptotic behavior or relies on various modeling assumptions
(e.g., \citeauthor{Robins94Estimation}, \citeyear
{Robins94Estimation}; \citeauthor{LD04}, \citeyear{LD04}; 
Kang and Schafer, \citeyear{KangSc07}).
%In contrast, o
Our analysis is nonasymptotic and makes no such assumptions.

Several papers in machine learning have used ideas related to the basic
technique discussed here, although not with the same language. For
\emph{benign bandits}, \citet{HK} construct algorithms which
use reward estimators to improve regret bounds when the
variance of actual rewards is small. Similarly,
the Offset Tree algorithm (\citeauthor{Beygelzimer09Offset}, \citeyear
{Beygelzimer09Offset}) can be thought of
as using a crude reward estimate for the ``offset.'' The
algorithms and estimators described here are substantially more
sophisticated.

Our nonstationary policy evaluation builds on the rejection sampling approach,
which has been previously shown to be effective (\citeauthor{LCLW11},
\citeyear{LCLW11}).
Relative to this earlier work, our nonstationary results take
advantage of the doubly robust technique and a carefully introduced
bias/variance tradeoff to obtain an empirical order-of-magnitude
improvement in evaluation quality.

%s2.2 #&#
\subsection{Dynamic Treatment Regimes} \label{sec:related-dtr}

Contextual bandit problems are closely related to dynamic treatment
regime (DTR) estimation/opti\-mization in medical research. A DTR is a
set of (possibly randomized) rules that specify what treatment to
choose, given current characteristics (including past treatment history
and outcomes) of a patient. In the terminology of the present paper,
the patient's current characteristics are contextual information, a
treatment is an action, and a DTR is a policy. Similar to contextual
bandits, the quantity of interest in DTR can be expressed by a numeric
reward signal related to the clinical outcome of a treatment. We
comment on similarities and differences between DTR and contextual
bandits in more detail in later sections of the paper, where we define
our setting more formally. Here, we make a few higher-level remarks.

Due to ethical concerns, research in DTR is often performed with
observational data rather than on patients. This corresponds to the
offline version of contextual bandits, which only has access to past
data but no ability to gather new data. Causal inference techniques
have been studied to estimate the mean response of a given DTR (e.g.,
\citeauthor{Robins86New}, \citeyear{Robins86New}; \citeauthor
{Murphy01Marginal}, \citeyear{Murphy01Marginal}), and to optimize DTR (e.g.,
\citeauthor{Murphy03Optimal}, \citeyear{Murphy03Optimal}; \citeauthor
{Orellana10Dynamic}, \citeyear{Orellana10Dynamic}). These two problems
correspond to evaluation and optimization of policies in the present paper.

%However, an optimal DTR in general aims at maximizing the total
%multi-step rewards accumulated over time on individual patients, so
%has to consider long-term effects of a treatment on a patient's health
%in the future.
In DTR, however, a treatment typically exhibits a long-term effect on a
patient's future ``state,'' while in contextual bandits the contexts
are drawn IID with no dependence on actions taken previously. Such a
difference turns out to enable statistically more efficient estimators,
which will be explained in greater detail in Section~\ref
{sec:dtr-nonstationary}.

%In contrast, in contextual bandits contexts are drawn IID over
%different rounds without being affected by previous actions taken. In
%this sense, optimal policies in a contextual bandit setting can be
%myopic by maximizing rewards received in every individual step.
Despite these differences, as we will see later, contextual bandits and
DTR share many similarities, and in some cases are almost identical.
For example, analogous to the results introduced in this paper, doubly
robust estimators have been applied to DTR estimation (Murphy, van~der Laan and Robins, \citeyear{Murphy01Marginal}), and also used as a
subroutine for optimization in a
family of parameterized policies (\citeauthor{Zhang12Robust},
\citeyear{Zhang12Robust}).
%In the following sections, the relation and differences will be
%explained.
The connection suggests a broader applicability of DTR techniques
beyond the medical domain, for instance, to the Internet-motivated
problems studied in this paper.

%s3 #&#
\section{Evaluation of Stationary Policies} \label{sec:stationary}

%s3.1 #&#
\subsection{Problem Definition} \label{sec:definition}

We are interested in the \emph{contextual bandit} setting where on each
round:
\begin{enumerate}[3.]
\item[1.] A vector of covariates (or a \emph{context}) $x\in\X$ is revealed.
\item[2.] An action (or \emph{arm}) $a$ is chosen from a given set
$\A$.
\item[3.] A reward $r\in[0,1]$ for the action $a$ is revealed, but the
rewards of other actions are not. The reward may
depend stochastically on $x$ and $a$.
\end{enumerate}

We assume that contexts are chosen IID from an unknown distribution $D(x)$,
the actions are chosen from a finite (and typically not too large)
action set $\A$,
and the distribution over rewards $D(r\vert a,x)$ does not change over time
(but is unknown).

The input data consists of a finite stream of triples $(x_k,a_k,r_k)$
indexed by $k=1,2,\ldots,n$. We
assume that the actions $a_k$ are generated by some past
(possibly nonstationary) policy, which we refer to as the \emph
{exploration policy}. The \emph{exploration history}
up to round $k$ is denoted
\[
z_k=(x_1,a_1,r_1,
\ldots,x_k,a_k,r_k) .
\]
%
%and an infinite exploration history $\Parens{x_k,a_k,r_k}_{k=1}^

Histories %$z$
are viewed as samples from a probability measure $\mu$. Our
assumptions about data generation
then translate into the assumption about factoring of $\mu$ as
\begin{eqnarray*}
&&\mu(x_k,a_k,r_k\vert z_{k-1})\\
&&\quad =
D(x_k) \mu(a_k\vert x_k,z_{k-1})
D(r_k\vert x_k,a_k) ,
\end{eqnarray*}
for any $k$. Note that apart
from the unknown distribution $D$, the only degree
of freedom above is $\mu(a_k\vert x_k,z_{k-1})$, that is, the unknown
exploration
policy.

When $z_{k-1}$
is clear from the context, we use a shorthand $\mu_k$ for the conditional
distribution over the $k$th triple
\[
\mu_k(x,a,r) = \mu(x_k=x, a_k=a,
r_k=r\vert z_{k-1} ) .
\]
We also write $\Pr_k^\mu$ and $\E^\mu_k$ for $\Pr_\mu[\,\cdot\,
\vert
z_{k-1}]$ and
$\E_{\mu}[\,\cdot\,\vert z_{k-1}]$.

Given input data $z_n$, we study the \emph{stationary policy
evaluation} problem.
A stationary randomized policy $\nu$ is described by a conditional
distribution $\nu(a\vert x)$ of
choosing an action on each context. The goal is to use the history
$z_n$ to estimate the \emph{value} of
$\nu$, namely, the expected reward obtained by following $\nu$:
\[
V(\nu) = \E_{x\sim D}\E_{a\sim\nu(\,\cdot\,\vert x)}\E_{r\sim
D(\,\cdot\,\vert x,a)}[r] .
\]
In content recommendation on Web portals, for example, $V(\nu)$
measures the average click probability per user visit, one of the major
metrics with critical business importance.

In order to have unbiased policy evaluation, we make a standard
assumption that if $\nu(a\vert x)>0$
then $\mu_k(a\vert x)>0$ for all $k$ (and all possible histories
$z_{k-1}$). This clearly holds
for instance if $\mu_k(a\vert x)>0$ for all $a$.
Since $\nu$ is fixed in our paper, we will write $V$ for $V(\nu)$.
To simplify notation, we extend the conditional distribution $\nu$ to
a distribution over triples $(x,a,r)$
\[
\nu(x,a,r) = D(x)\nu(a\vert x)D(r\vert a,x)
\]
and hence $V=\E_{\nu}[r]$.

The problem of stationary policy evaluation, defined above, is slightly
more general than DTR analysis in a typical cross-sectional
observational study, where the exploration policy (known as ``treatment
mechanism'' in the DTR literature) is stationary; that is, the
conditional distribution $\mu(a_k|x_k,z_{k-1})$ is independent of
$z_{k-1}$ and identical across all $k$, that is, $\mu_k=\mu_1$ for
all~$k$.
% and equals some fixed distribution $\bar{\mu}_{A|X}$ for all $k$.

%s3.2 #&#
\subsection{Existing Approaches}

The key challenge in estimating policy value in contextual bandits is
that rewards are \emph{partially} observable: in each round, only the
reward for the chosen action is revealed; we do not know what the
reward would have been if we chose a different action. Hence, the data
collected in a contextual bandit process cannot be used directly to
estimate a new policy's value: if in a context $x$ the new policy
selects an action $a'$ different from the action $a$ chosen during data
collection, we simply do \emph{not} have the reward signal for $a'$.

There are two common solutions for overcoming this limitation
(see, e.g., \citeauthor{DRAds}, \citeyear{DRAds}, for an introduction
to these solutions).
The first, called the \emph{direct method} (DM), forms an estimate
$\hr(x,a)$ of the expected reward conditioned on the context
\emph{and} action. The policy value is then estimated by
\[
\hV_\DM= \frac{1}{n}\sum_{k=1}^n
\sum_{a\in\A} \nu(a\vert x_k)
\hr(x_k,a) .
\]
Clearly, if $\hr(x,a)$ is a good approximation of the true expected
reward $\E_{D}[r\vert x,a]$,
then the DM estimate is close to $V$.
%
%Also, if $\hr$ is unbiased, $\smash{\hat{V}}\vphantom{V}_\DM$
%is an unbiased
%estimate of $V$.
%
A problem with this method is that the estimate
$\hr$ is typically formed without the knowledge of $\nu$, and hence
might focus
on approximating expected reward
in the areas that are irrelevant for $\nu$ and not sufficiently in the
areas that are
important for $\nu$ (see, e.g., the analysis of \citeauthor
{Beygelzimer09Offset}, \citeyear{Beygelzimer09Offset}).

The second approach, called \emph{inverse propensity score} (IPS), is
typically less prone to problems with bias. Instead of
approximating the reward, IPS forms an approximation $\hmu_k(a\vert x)$
of $\mu_k(a\vert x)$, and uses this estimate to correct for the shift in
action proportions between the exploration policy and the new policy:
\[
\hV_\IPS= \frac{1}{n}\sum_{k=1}^n
\frac{\nu(a_k\vert x_k)}{\hmu
_k(a_k\vert x_k)}\cdot r_k .
\]
If $\hmu_k(a\vert x)\approx\mu_k(a\vert x)$, then the IPS estimate
above will be,
approximately, an unbiased estimate of $V$. Since we typically have a
good (or even
accurate) understanding of the data-collection policy, it is often
easier to obtain good estimates $\hmu_k$, and thus the IPS estimator
is in
practice less susceptible to problems with bias compared with the
direct method. However, IPS typically has a much larger variance, due
to the increased range of the random variable $\nu(a_k\vert x_k)/\hmu
_k(a_k\vert x_k)$. The issue becomes
more severe when $\hmu_k(a_k\vert x_k)$ gets smaller in high
probability areas under $\nu$. Our approach
alleviates the large variance problem of IPS by taking advantage of
the estimate $\hr$ used by the direct method.
%
% MD: maybe say something about bias variance control through \hmu_k?

%s3.3 #&#
\subsection{Doubly Robust Estimator}

Doubly robust estimators take advantage of both the estimate of the
expected reward $\hr$
and the estimate of action probabilities $\hmu_k(a\vert x)$.
A similar idea has been suggested earlier by a number of authors for
different estimation problems (\citeauthor{CasselSaWr76}, \citeyear
{CasselSaWr76}; Rotnitzky and Robins, \citeyear
{Rotnitzky95Semiparametric};
\citeauthor{RR95}, \citeyear{RR95}; Murphy, van~der Laan and Robins,
\citeyear{Murphy01Marginal}; \citeauthor{Robins98Marginal}, \citeyear
{Robins98Marginal}).
For the setting in this section, the estimator of \citet
{Murphy01Marginal} can be reduced to
%
%e3.1 #&#
%
\begin{eqnarray}
\label{eq:DR} \hspace*{20pt}\hV_\DR&=& \frac{1}{n}\sum
_{k=1}^n \biggl[\hat{r}(x_k,\nu)\\
&&\hphantom{\frac{1}{n}\sum
_{k=1}^n \biggl[}{} +
\frac{\nu(a_k\vert x_k)}{\hmu_k(a_k\vert
x_k)} \cdot\bigl(r_k-\hr(x_k,a_k)\bigr)
\biggr],\nonumber
\end{eqnarray}
where
\[
\hat{r}(x,\nu) = \sum_{a\in\A} \nu(a\vert x)\hr(x,a)
\]
is the estimate of $\E_\nu[r\vert x]$ derived from $\hr$. Informally,
the doubly robust estimator uses $\hr$ as a baseline and if there is
data available,
a correction is applied. We will see that our estimator is unbiased if
\emph{at least one}
of the estimators, $\hr$ and $\hmu_k$, is accurate, hence the name
\emph{doubly robust}.

In practice, quite often neither $\E_D[r\vert x,a]$ or $\mu_k$ is accurate.
%Indeed, accurate estimation of $\E_D[r\given x,a]$ often requires
%precise knowledge of the reward generation process that is seldom
%available (e.g., how a user clicks on an advertisement).
It should be noted that, although $\mu_k$ tends to be much easier to
estimate than $\E_D[r\vert x,a]$ in applications that motivate this
study, it is rare to be able to get a perfect estimator, due to
engineering constraints in complex applications like Web search and
Internet advertising.
Thus, a basic question is: How does the estimator $\hV_\DR$ perform
as the estimates
$\hr$ and $\hmu_k$ deviate from the truth?
%We first analyze the bias of the estimator. In the next section, we
%analyze its variance.
The following section analyzes bias and variance of the DR estimator as
a function
of errors in $\hr$ and $\hmu_k$. Note that our DR estimator
encompasses DM and IPS as
special cases (by respectively setting $\hmu_k\equiv\infty$ and $\hr
\equiv0$),
so our analysis also encompasses DM and IPS.
\setcounter{footnote}{1}
%s3.4 #&#
\subsection{Analysis}
\label{sec:dr-analysis}

We assume that $\hr(x,a)\in[0,1]$ and $\hmu_k(a\vert x)\in(0,\infty
]$, but in general $\hmu_k$ does not need to represent conditional
probabilities (our notation is only meant to indicate that $\hmu_k$
estimates $\mu_k$, but no probabilistic structure). In general, we
allow $\hr$ and $\hmu_k$ to be random variables, as long as they
satisfy the following independence assumptions:
\begin{itemize}
\item$\hr$ is independent of $z_n$.
\item$\hmu_k$ is conditionally independent of $\{(x_\ell,a_\ell
,r_\ell)\}_{\ell\ge k}$, conditioned on $z_{k-1}$.
\end{itemize}
The first assumption means that $\hr$ can be assumed fixed and
determined before we see the input data $z_n$, for example,
by initially splitting the input dataset and using the first part to
obtain $\hr$ and the
second part to evaluate the policy. In our analysis, we condition on
$\hr$ and ignore any randomness in its choice.

The second assumption means that $\hmu_k$ is not allowed to depend on
future. A simple way to satisfy this assumption is to
split the dataset to form an estimator (and potentially also include
data $z_{k-1}$).
If we have some control over
the exploration process, we might also have access to ``perfect
logging'', that is,
recorded probabilities $\mu_k(a_k\vert x_k)$.
%Since we only need to apply the estimator $\hmu_k$ to obtain $
With perfect logging, we can achieve
$\hmu_k=\mu_k$, respecting our assumptions.\footnote{%
As we will see later in the paper, in order
to reduce the variance of the estimator it might still be advantageous
to use a slightly inflated estimator, for example, $\hmu_k=c\mu_k$
for $c>1$, or $\hmu_k(a\vert x)=\max\{c,\mu_k(a\vert x)\}$
for some $c>0$.}

Analogous to $\hr(x,a)$, we define the population quantity $r^*(x,a)$
\[
r^*(x,a) =\E_D[r\vert x,a] ,
\]
and define $r^*(x,\nu)$ similarly to $\hat{r}(x,\nu)$:
\[
r^*(x,\nu) =\E_\nu[r\vert x] .
\]

Let $\Delta(x,a)$ and $\q_k(x,a)$ denote, respectively,
the \textit{additive} error of $\hr$ and the \textit{multiplicative}
error of $\hmu_k$:
\begin{eqnarray*}
\Delta(x,a)&=&\hr(x,a) - r^*(x,a),
\\
\q_k(x,a) &=& \mu_k(a\vert x) / \hmu_k(a
\vert x) .
\end{eqnarray*}
We assume that for some $M\ge0$, with probability one under $\mu$:
\[
\nu(a_k\vert x_k)/\hmu_k(a_k
\vert x_k)\le M
\]
which can always be satisfied by enforcing $\hmu_k \geq1/M$.

To bound the error of $\hV_\DR$, we first analyze a single term:
\[
\hV_{k}=\hat{r}(x_k,\nu) + \frac{\nu(a_k\vert x_k)}{\hmu
_k(a_k\vert x_k)} \cdot
\bigl(r_k-\hr(x_k,a_k) \bigr) .
\]
We bound its range, bias, and conditional variance as follows (for
proofs, see
Appendix~\ref{app:proofs1}):
%
%le3.1 #&#
%
\begin{lemma}
\label{LEMMA:RANGE}
The range of $\hV_{k}$ is bounded as
\[
\vert\hV_{k} \vert\le1 + M .
\]
\end{lemma}
%
%le3.2 #&#
%
\begin{lemma}
\label{lemma:exp}
The expectation of the term $\hV_{k}$ is
\[
\E^\mu_k[\hV_{k}] = \EE_{(x,a)\sim\nu}
\bigl[r^*(x,a) + \bigl(1-\q_k(x,a) \bigr)\Delta(x,a) \bigr] .
\]
\end{lemma}
%
%le3.3 #&#
%
\begin{lemma}
\label{LEMMA:VAR}
The variance of the term $\hV_{k}$ can be decomposed and bounded as follows:
\renewcommand{\theequation}{\roman{equation}}{
\setcounter{equation}{0}
%e3.2 #&#
%e3.3 #&#
%
\begin{eqnarray}
\hspace*{10pt} &&\Var^\mu_k[\hV_{k}]\\
&&\quad=
\VV_{x\sim D} \Bigl[\EE_{a\sim\nu(\,\cdot\,
\vert x)} \bigl[r^*(x,a)\nonumber\\
&&\hphantom{\quad=\VV_{x\sim D} \bigl[\EE_{a\sim\nu(\,\cdot\,
\vert x)} \bigl[}{}+ \bigl(1-
\q_k(x,a) \bigr)\nonumber\\
&&\hspace*{114pt} {}\cdot\Delta(x,a) \bigr] \Bigr]
\nonumber
\\
&&\qquad {} - \EE_{x\sim D} \Bigl[\EE_{a\sim\nu(\,\cdot\,\vert
x)} \bigl[\q
_k(x,a)\Delta(x,a) \bigr]^2 \Bigr]
\nonumber
\\
&&\qquad {} + \EE_{(x,a)\sim\nu} \biggl[\frac{\nu(a\vert
x)}{\hmu_k(a\vert x)}\nonumber\\
&&\hphantom{\qquad{} + \EE_{(x,a)\sim\nu} \biggl[}{} \cdot
\q_k(x,a)\cdot\VV_{r\sim D(\,\cdot\,\vert x,a)}[r] \biggr]
\nonumber
\\
&&\qquad {} + \EE_{(x,a)\sim\nu} \biggl[\frac{\nu(a\vert
x)}{\hmu_k(a\vert x)}\nonumber\\
&&\hphantom{\qquad{} + \EE_{(x,a)\sim\nu} \biggl[}{}\cdot
\q_k(x,a)\Delta(x,a)^2 \biggr].\nonumber
\\
&& \Var^\mu_k[\hV_{k}]\\
&&\quad\le
\VV_{x\sim D} \bigl[r^*(x,\nu) \bigr]\nonumber\\
&&\qquad{} + 2\EE_{(x,a)\sim\nu} \bigl[ \bigl\vert
\bigl(1-\q_k(x,a) \bigr)\Delta(x,a) \bigr\vert\bigr]
\nonumber
\\
&&\qquad {} + M\EE_{(x,a)\sim\nu} \Bigl[\q_k(x,a)\nonumber\\
&&\hphantom{\qquad{} + M\EE_{(x,a)\sim\nu} \bigl[}{}\cdot\EE
_{r\sim D(\,\cdot\,\vert
x,a)} \bigl[ \bigl(r-\hr(x,a) \bigr)^2 \bigr] \Bigr] .\nonumber
\end{eqnarray}
}
\end{lemma}

The range of $\hV_{k}$ is controlled by the worst-case ratio $\nu
(a_k\vert x_k)/\hmu_k(a_k\vert x_k)$.
The bias of $\hV_{k}$ gets smaller as $\Delta$ and $\q_k$ become
more accurate, that is, as $\Delta\approx0$
and $\q_k\approx1$. The expression for variance is more complicated.
Lemma~\ref{LEMMA:VAR}(i) lists four
terms. The first term represents the variance component due to the
randomness over $x$. The
second term can contribute to the decrease in the variance.
%(for example if $\q_k\approx1$, and $\Delta$ tends to have the same
%sign).
The final two terms represent
the penalty due to the importance weighting.
The third term scales with the conditional variance of rewards (given
contexts and actions),
and it vanishes if rewards are deterministic. The fourth term scales
with the magnitude of $\Delta$,
and it captures the potential improvement due to the use of a good
estimator $\hr$.
%by using DR as opposed to IPS.

The upper bound on the variance [Lemma~\ref{LEMMA:VAR}(ii)] is easier
to interpret. The first term is the
variance of the estimated variable over $x$. The second term measures
the quality of the estimators $\hmu_k$
and $\hr$---it equals zero if either of them is perfect (or if the
union of regions where they are perfect
covers the support of $\nu$ over $x$ and $a$). The final term
represents the importance weighting penalty. It
vanishes if we do not apply importance weighting (i.e., $\hmu_k\equiv
\infty$ and $\q_k\equiv0$). With nonzero
$\q_k$, this term decreases with a better quality of $\hr$---but it
does not disappear even if $\hr$ is perfect
(unless the rewards are deterministic).

%s3.4.1 #&#
\subsubsection{Bias analysis}

Lemma~\ref{lemma:exp} immediately yields a bound on the bias of the
doubly robust
estimator, as stated in the following theorem. The special case for
stationary policies (second part of the theorem) has been shown by
\citet{Vansteelandt10Model}.
%
%th3.4 #&#
%
\begin{theorem}
\label{thm:DR:bias}
Let $\Delta$ and $\q_k$ be defined as above. Then the bias of the
doubly robust
estimator is
\begin{eqnarray*}
&&\bigl\vert\E_\mu[\hV_\DR] - V \bigr\vert\\
&&\quad= \frac{1}{n}\Biggl
\llvert\E_\mu\Biggl[\sum_{k=1}^n
\EE_{(x,a)\sim
\nu} \bigl[ \bigl(1-\q_k(x,a) \bigr)\Delta(x,a) \bigr]
\Biggr] \Biggr\rrvert.
\end{eqnarray*}
If the exploration policy $\mu$ and the estimator $\hmu_k$ are stationary
(i.e., $\mu_k=\mu_1$ and $\hmu_k=\hmu_1$ for all $k$), the
expression simplifies to
%
%e3.4 #&#
%
\[
\nonumber
\bigl\vert\E_\mu[\hV_\DR] - V \bigr\vert=
\bigl\vert\E_\nu\bigl[ \bigl(1-\q_1(x,a) \bigr)\Delta(x,a)
\bigr] \bigr\vert.
\]
\end{theorem}
\begin{pf}
The theorem follows immediately from Lemma~\ref{lemma:exp}.
\end{pf}
In contrast, we have (for simplicity, assuming stationarity of the
exploration policy and its estimate)
\begin{eqnarray*}
\bigl\vert\E_\mu[\hV_\DM] - V \bigr\vert&=& \bigl\vert\E_\nu
\bigl[\Delta(x,a) \bigr] \bigr\vert,
\\
\bigl\vert\E_\mu[\hV_\IPS] - V \bigr\vert&=& \bigl\vert\E
_\nu
\bigl[r^*(x,a) \bigl(1-\q_1(x,a) \bigr) \bigr] \bigr\vert,
\end{eqnarray*}
where the first equality is based on the observation that DM
is a special case of DR with $\hmu_k(a\vert x)\equiv\infty$ (and
hence $\q_k\equiv0$), and
the second equality is based on the observation that IPS is a
special case of DR with $\hr(x,a)\equiv0$ (and hence $\Delta\equiv r^*$).

In general, neither of the estimators dominates the others. However,
if \emph{either} $\Delta\approx0$, \emph{or} $\q_k\approx1$, the
expected
value of the doubly robust
estimator will be close to the true value, whereas DM requires
$\Delta\approx0$ and IPS requires $\q_k\approx1$. Also, if %$\Delta
$\Vert\q_k-1\Vert_{p,\nu}\ll1$ [for a suitable $L_p(\nu)$
norm], we expect that DR will
outperform DM. Similarly, if $\q_k\approx1$ but $\Vert\Delta\Vert
_{p,\nu}\ll\Vert r^*\Vert_{p,\nu}$,
we expect that DR will outperform IPS.
Thus, DR can effectively take advantage of
both sources of information to lower the bias.

%s3.4.2 #&#
\subsubsection{Variance analysis}

We argued that the expected value of
$\hV_\DR$ compares favorably with IPS and DM. We next
look at the variance of DR. Since large-deviation bounds have a
primary dependence on variance; a lower variance implies a faster
convergence rate. To contrast DR with IPS and DM, we study a
simpler setting
with a stationary exploration policy, and \emph{deterministic
target policy} $\nu$, that is, $\nu(\,\cdot\,\vert x)$ puts all the
probability on a single action. In the next section, we revisit the
fully general setting and derive a finite-sample bound
on the error of DR.
%
%th3.5 #&#
%
\begin{theorem}
\label{thm:DR:var}
Let $\Delta$ and $\q_k$ be defined as above. If exploration policy
$\mu$
and the estimator $\hmu_k$ are stationary, and the target policy $\nu$
is deterministic, then the variance of the doubly
robust estimator is
\begin{eqnarray*}
&&\hspace*{-5pt}\Var_\mu[\hV_\DR] \\
&&\hspace*{-5pt}\quad= \frac{1}{n} \biggl(
\VV_{(x,a)\sim\nu} \bigl[r^*(x,a)\\
&&\hspace*{-5pt}\hphantom{\quad= \frac{1}{n} \biggl(\VV
_{(x,a)\sim\nu} \bigl[}{}+ \bigl(1-\q_1(x,a) \bigr)\Delta
(x,a) \bigr]
\\
&&\hspace*{-5pt}\hphantom{\quad= \frac{1}{n} \biggl(}{}+ \EE
_{(x,a)\sim\nu} \biggl[\frac{1}{\hmu_1(a\vert x)} \cdot\q
_1(x,a)\cdot
\VV_{r\sim D(\,\cdot\,\vert x,a)}[r] \biggr]\\
&&\hspace*{-5pt}\hphantom{\quad= \frac{1}{n} \biggl(}{} + \EE
_{(x,a)\sim\nu} \biggl[\frac{1-\mu_1(a\vert x)}{\hmu_1(a\vert
x)} \cdot
\q_1(x,a)\Delta(x,a)^2 \biggr] \biggr) .
\end{eqnarray*}
\end{theorem}
\begin{pf}
The theorem follows immediately from Lemma~\ref{LEMMA:VAR}(i).
\end{pf}

The variance can be decomposed into three terms.
The first term accounts for the randomness in $x$ (note that $a$ is
deterministic given $x$).
The other two terms can be viewed as the importance weighting penalty.
These two terms
disappear in DM, which does not use
rewards $r_k$.
The second term accounts for randomness in rewards and disappears when rewards
are deterministic functions of $x$ and $a$. However, the last term stays,
accounting for the disagreement between actions taken by $\nu$ and
$\mu_1$.

Similar expressions can be derived for the DM and IPS estimators. Since
IPS is a special case of DR with $\hr\equiv0$,
we obtain the following equation:
\begin{eqnarray*}
&&\hspace*{-5pt}\Var_\mu[\hV_\IPS]\\
&&\hspace*{-5pt}\quad = \frac{1}{n} \biggl(
\VV_{(x,a)\sim\nu} \bigl[\q_1(x,a)r^*(x,a) \bigr]
\\
&&\hspace*{-5pt}\hphantom{\quad = \frac{1}{n} \biggl(}{}+ \EE
_{(x,a)\sim\nu} \biggl[\frac{1}{\hmu_1(a\vert x)} \cdot\q
_1(x,a)\cdot
\VV_{r\sim D(\,\cdot\,\vert x,a)}[r] \biggr] \\
&&\hspace*{-5pt}\hphantom{\quad = \frac{1}{n} \biggl(}{}+ \EE
_{(x,a)\sim\nu} \biggl[\frac{1-\mu_1(a\vert x)}{\hmu_1(a\vert
x)} \cdot
\q_1(x,a)r^*(x,a)^2 \biggr] \biggr) .
\end{eqnarray*}
The first term will
be of similar magnitude as the corresponding term of the DR estimator,
provided that $\q_1\approx1$. The second term is identical to the DR
estimator.
However, the third term
can be much larger for IPS if $\mu_1(a\vert x)\ll1$
and $\vert\Delta(x,a)\vert$ is
smaller than $r^*(x,a)$ for the actions chosen by $\nu$.

In contrast, for the direct method, which is a special case of DR with
$\hmu_k\equiv\infty$, the following variance is obtained immediately:
\[
\Var_\mu[\hV_\DM] = \frac{1}{n}
\VV_{(x,a)\sim\nu} \bigl[r^*(x,a)+\Delta(x,a) \bigr] .
\]
Thus, the variance of the direct method does not have terms depending
either on the exploration policy or the randomness in the rewards. This fact
usually suffices to ensure that its variance is significantly lower
than that
of DR or IPS. However, as mentioned in the previous
section, when we can estimate $\mu_k$ reasonably well (namely, $\q
_k\approx1$), the bias of the direct method is typically much larger,
leading to larger errors in estimating policy values.

%s3.4.3 #&#
\subsubsection{Finite-sample error bound}
\label{sec:error}

By combining bias and variance bounds, we now work out a specific
finite-sample bound on the
error of the estimator $\hV_\DR$. While such an error bound could
be used as a conservative confidence bound, we expect it to be too
loose in most settings (as is typical for finite-sample bounds). Instead,
our main intention is to explicitly highlight how the errors of estimators
$\hr$ and $\hmu_k$ contribute to the final error.

To begin, we first quantify magnitudes of the additive error $\Delta
=\hr-r^*$
of the estimator $\hr$, and the relative error $\vert1-\q_k\vert
=\vert\hmu_k-\mu_k\vert/\hmu_k$
of the estimator~$\hmu_k$:
%
%as3.6 #&#
%
\begin{assumption}
\label{assume:1}
Assume there exist $\delta_\Delta,\delta_\q\ge0$ such that
\[
\EE_{(x,a)\sim\nu} \bigl[\bigl\llvert\Delta(x,a)\bigr\rrvert
\bigr] \le
\delta_\Delta,
\]
and with probability one under $\mu$:
\[
\bigl\vert1-\q_k(x,a)\bigr\vert\le\delta_\q\quad\mbox{for all $k$.}
\]
\end{assumption}

Recall that $\nu/\hmu_k\le M$.
In addition, our analysis depends on the magnitude of the ratio $\q
_k=\mu_k/\hmu_k$
and a term that captures both the variance of the rewards and the error
of $\hr$.
%
%as3.7 #&#
%
\begin{assumption}
\label{assume:2}
Assume there exist $e_{\hr},\qmax\ge0$ such that with probability
one under $\mu$, for all $k$:
\begin{eqnarray*}
&&\EE_{(x,a)\sim\nu} \Bigl[\EE_{r\sim D(\,\cdot\,\vert x,a)}
\bigl[ \bigl(\hr(x,a)-r
\bigr)^2 \bigr] \Bigr] \le e_{\hr},
\\
&& \q_k(x,a) % =
% \mu_k(a\given x)/\hmu_k(a\given x)
\le\qmax\quad\mbox{for all $x,a$.}
\end{eqnarray*}
\end{assumption}

With the assumptions above, we can now bound the bias and variance of a
single term $\hV_{k}$. As
in the previous sections, the bias decreases with the quality of $\hr$
and $\hmu_k$, and the
variance increases with the variance of the rewards
and with the magnitudes of the ratios $\nu/\hmu_k\le M$, $\mu_k/\hmu
_k\le\qmax$. The
analysis below for instance captures the bias-variance tradeoff of
using $\hmu_k\approx c\mu_k$ for some $c>1$: such a strategy can lead
to a lower variance
(by lowering $M$ and $\qmax$) but incurs some additional bias that is
controlled by the quality of~$\hr$.
%
%le3.8 #&#
%
\begin{lemma}
\label{lemma:bound:stationary}
Under Assumptions \ref{assume:1}--\ref{assume:2},
with probability one under $\mu$, for all $k$:
\begin{eqnarray*}
\bigl\llvert\E^\mu_k[\hV_{k}]-V\bigr
\rrvert&\le&\delta_\q\delta_\Delta,
\\
\Var^\mu_k[\hV_{k}] &\le& \Var_{x\sim D}
\bigl[r^*(x,\nu)\bigr] +2\delta_\q\delta_\Delta+ M\qmax
e_{\hr} .
\end{eqnarray*}
\end{lemma}
\begin{pf}
%The range bound is just Lemma~\ref{lemma:range}.
The bias and variance
bound follow from Lemma~\ref{lemma:exp}
and Lemma~\ref{LEMMA:VAR}(ii), respectively, by H\"ol\-der's inequality.
\end{pf}

Using the above lemma and Freedman's inequality yields the following theorem.
%
%th3.9 #&#
%
\begin{theorem}
\label{thm:bound:stationary}
Under Assumptions \ref{assume:1}--\ref{assume:2},
with probability at least $1-\delta$,
{\fontsize{10.6}{10.9}\selectfont{
\begin{eqnarray*}
&&\hspace*{-5pt}\vert\hV_\DR- V \vert\\
&&\hspace*{-6pt}\quad\le\delta_\q\delta_\Delta\\
&&\hspace*{-7pt}\qquad{}+ 2\max\biggl\{ \frac{(1+M)\ln(2/\delta)}{n},
\\
&&\hspace*{-7pt}\qquad\sqrt{\frac{ (\Var_{x\sim D}[r^*(x,\nu)]
+2\delta_\q\delta
_\Delta+ M\qmax e_{\hr} )\ln(2/\delta)}{n}} \biggr\} .
\end{eqnarray*}
}}
\end{theorem}
\begin{pf}
The proof follows by Freedman's inequality (Theorem~\ref{thm:freedman}
in Appendix~\ref{app:freedman}), applied to random variables $\hV
_{k}$, whose range and variance are bounded using Lemmas \ref
{LEMMA:RANGE} and \ref{lemma:bound:stationary}.
\end{pf}

The theorem is a finite-sample error bound that holds for all sample
size $n$, and in the limit the error converges to $\delta_\q\delta
_\Delta$. As we mentioned, this result gives a confidence interval for
the doubly-robust estimate $\hV_\DR$ for any finite sample $n$. Other
authors have used asymptotic theory to derive confidence intervals for
policy evaluation by showing that the estimator is asymptotically
normal (e.g., \citeauthor{Murphy01Marginal}, \citeyear
{Murphy01Marginal}; Zhang\break et~al., \citeyear{Zhang12Robust}).
When using asymptotic confidence bounds, it can be difficult to know a
priori whether the asymptotic distribution has been reached,\break
whereas our bound applies to all finite sample sizes.
Although our bound may be conservative for small sample sizes, it
provides a ``safe'' nonasymptotic confidence interval. In certain
applications like those on the Internet, the sample size is usually
large enough for this kind of nonasymptotic confidence bound to be
almost as small as its asymptotic value (the term $\delta_\rho\delta
_\Delta$ in Theorem~\ref{thm:bound:stationary}), as demonstrated by
\citet{Bottou13Counterfactual} for online advertising.
%
%t1 #&#
%
\begin{table*}[t]
\tablewidth=\textwidth
\tabcolsep=0pt
\caption{Characteristics of benchmark datasets used in
Section~\protect\ref{sec:class}}
\label{tab:class-data}
\begin{tabular*}{\textwidth}{@{\extracolsep{\fill}}lccccccccc@{}}
\hline
\textbf{Dataset} & \textbf{Ecoli} & \textbf{Glass} & \textbf
{Letter} & \textbf{Optdigits} &
\textbf{Page-blocks} & \textbf{Pendigits}
& \textbf{Satimage} & \textbf{Vehicle} & \textbf{Yeast} \\
\hline
Classes ($K$) & \hphantom{33}8 & \hphantom{21}6 & \hphantom{20\mbox
{,}0}26 & \hphantom{56}10 &
\hphantom{547}5 & \hphantom{10\mbox{,}9}10 & \hphantom{643}6 &
\hphantom{84}4 & \hphantom{14}10 \\
Sample size & 336 & 214 & 20\mbox{,}000 & 5620
& 5473 & 10\mbox{,}992 & 6435 & 846
& 1484 \\
\hline
\end{tabular*}
\end{table*}

%
%f1 #&#
%
\begin{figure*}[b]

\includegraphics{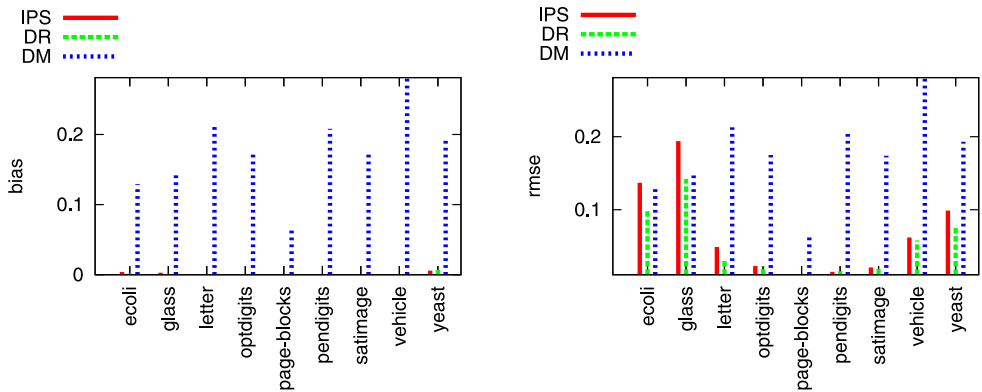}

\caption{Comparison of \bias\ (left) and \rmse\ (right) of the three
estimators of classification error on partial feedback classification
data.} \label{fig:class-eval}
\end{figure*}

Note that Assumptions \ref{assume:1}--\ref{assume:2} rely on bounds
of $\vert1-\q_k\vert$ and $\q_k$ which have
to hold with probability one. In Appendix~\ref{app:error}, we replace
these bounds
with moment bounds, and present analogs of
Lemma~\ref{lemma:bound:stationary} and Theorem~\ref{thm:bound:stationary}.

%s4 #&#
\section{Experiments: the Stationary Case}
\label{sec:experiment1}

This section provides empirical evidence for the effectiveness
of the DR estimator compared to IPS and DM.
We study these estimators on several real-world datasets.
First, we use public benchmark datasets for multiclass classification
to construct contextual bandit data, on which we evaluate both
policy evaluation and policy optimization approaches.
Second, we use a proprietary dataset to model the pattern of user
visits to an Internet portal. We study covariate shift, which can
be formalized as a special case of policy evaluation.
Our third experiment uses another proprietary dataset
to model slotting of various types of search results on a webpage.

%s4.1 #&#
\subsection{Multiclass Classification with Partial Feedback}
\label{sec:class}

%We first reformulate multiclass classification tasks as contextual
%bandit problems for empirical study. This transformation allows us to
%compare IPS and DR using \emph{public} datasets for evaluating and
%minimizing classification error.

We begin with a description of how to turn a $K$-class classification
dataset into a $K$-armed contextual bandit dataset. Instead of rewards,
we will work with losses, specifically the 0$/$1-classification error.
The actions correspond to predicted classes. In the usual multiclass
classification, we can infer the loss of any action on training data
(since we know its correct label), so we call this a \emph{full
feedback} setting. On the other hand, in contextual bandits, we only
know the loss of the specific action that was taken by the exploration
policy, but of no other action, which we call a \emph{partial
feedback} setting. After choosing an exploration policy, our
transformation from full to partial feedback simply ``hides'' the
losses of actions that were not picked by the exploration policy.

This protocol gives us two benefits: we can carry out comparison using
\emph{public} multiclass classification datasets, which are more
common than contextual bandit datasets. Second, fully revealed data can
be used to obtain ground truth value of an arbitrary policy. Note that
the original data is real-world, but exploration and partial feedback
are simulated.

%s4.1.1 #&#
\subsubsection{Data generation}
\label{sec:class-data}

In a classification task, we assume data are drawn IID from a fixed
distribution: $(x,y)\sim D$, where $x\in\X$ is a real-valued
covariate vector and $y\in\{1,2,\ldots,K\}$ is a class label. A
typical goal is to find a classifier $\nu\dvtx\X\mapsto\{1,2,\ldots
,K\}$
minimizing the classification error:
\[
e(\nu)=\EE_{(x,y)\sim D} \bigl[\ind\bigl[\nu(x)\ne y\bigr] \bigr
] ,
\]
where $\ind[\,\cdot\,]$ is an indicator function, equal to $1$ if its
argument is true and $0$ otherwise.

The classifier $\nu$ can be viewed as a deterministic stationary
policy with the action set $\A=\{1,\ldots,K\}$ and the
\emph{loss function}
\[
l(y,a) = \ind[a\ne y] .
\]
Loss minimization is symmetric to the reward maximization (under
transformation $r=1-l$), but loss minimization
is more commonly used in classification setting, so we work with loss
here. Note that the distribution
$D(y\vert x)$ together with the definition of the loss above, induce
the conditional probability $D(l\vert x,a)$ in contextual bandits, and
minimizing the classification error coincides with policy optimization.

To construct partially labeled data in multiclass classification, it
remains to specify the exploration policy. We simulate stationary
exploration with $\mu_k(a\vert x)=\mu_1(a\vert x)=1/K$ for all $a$. Hence,
the original example $(x,y)$ is transformed into an example
$(x,a,l(y,a))$ for a randomly selected
action $a\sim\operatorname{uniform}(1,2,\ldots,K)$. We assume perfect
logging of the exploration policy
and use the estimator
$\hmu_k=\mu_k$. Below, we describe how we obtained an estimator $\hat
{l}(x,a)$
(the counterpart of $\hr$).

%Alternatively, we may turn the data point $(x,c)$ into a
%cost-sensitive classification example $(x,l_1,l_2,\ldots,l_k)$, where
%$l_a = \ind(a \ne c)$ is the loss for predicting $a$. Then, a
%classifier $\pi$ may be interpreted as an
%action-selection policy, and its classification error is exactly the
%policy's expected
%loss.\footnote{When considering classification problems, it is more
%natural to talk about minimizing
%classification errors. This loss minimization problem is symmetric to
%the reward maximization problem
%defined in Section~\ref{sec:definition}.}

%To construct a partially labeled dataset, exactly one loss component
%for each example is observed, following the %approach of
%$l_a$. The final data %are thus in the form of $(x,a,l_a)$,
%which is the form of data defined in Section~\ref{sec:definition}.
%Furthermore, $p(a\given x)\equiv1/k$ and is
%assumed to be known.

Table~\ref{tab:class-data} summarizes the benchmark problems adop\-ted
from the UCI repository (Asuncion and Newman, \citeyear
{Asuncion07Uci}).

%s4.1.2 #&#
\subsubsection{Policy evaluation}
\label{sec:class-eval}

We first investigate wheth\-er the DR technique indeed gives more
accurate estimates of the policy value (or classification error in our
context), compared to DM and IPS. For each dataset:
\begin{enumerate}[3.]
\item[1.] We randomly split data into training and evaluation sets of
(roughly) the same size;
\item[2.] On the training set, we keep full classification feedback of
form $(x,y)$
and train the direct loss minimization
(DLM) algorithm of \citet{McAllester11Direct}, based on gradient descent,
to obtain a classifier (see Appendix~\ref{app:dlm} for details). This
classifier
constitutes the policy $\nu$ whose value we estimate on evaluation data;
\item[3.] We compute the classification error on fully observed evaluation
data. This error is treated as the ground truth for comparing various
estimates;
\item[4.] Finally, we apply the transformation in Section~\ref
{sec:class-data} to the evaluation data to obtain a partially labeled
set (exploration history), from which DM, IPS and DR estimates are
computed. \label{step:class-eval}
\end{enumerate}

Both DM and DR require estimating the expected conditional loss for a
given $(x,a)$. We use a linear loss model: $\hat{l}(x,a)=w_a\cdot x$,
parameterized by $K$ weight vectors $\{w_a\}_{a\in\{1,\ldots,K\}}$,
and use least-squares ridge regression to fit $w_a$ based on the
training set.

Step~4 of the above protocol is repeated $500$
times, and the resulting
\bias\ and \rmse\  (root mean squared error) are reported in
Figure~\ref{fig:class-eval}.

As predicted by analysis, both IPS and DR are unbiased, since the
estimator $\hmu_k$ is perfect. In contrast, the linear loss model
fails to capture the classification error accurately, and as a result,
DM suffers a much larger bias.

While IPS and DR estimators are unbiased, it is apparent from the \rmse
\
plot that the DR estimator enjoys a lower variance, which translates
into a smaller \rmse. As we shall see next, this has a substantial
effect on the quality of policy optimization.

%s4.1.3 #&#
\subsubsection{Policy optimization}
\label{sec:class-opt}

This subsection deviates from much of the paper to study \emph{policy
optimization} rather than \emph{policy evaluation}. Given a space of
possible policies, policy optimization is a procedure that searches
this space for the policy with the highest value. Since policy values
are unknown, the optimization procedure requires access to exploration
data and uses a policy evaluator as a subroutine. Given the superiority
of DR over DM and IPS for policy evaluation (in previous subsection), a
natural question is whether a similar benefit can be translated into
policy optimization as well.
Since DM is significantly worse on all datasets, as indicated in
Figure~\ref{fig:class-eval}, we focus on the comparison between IPS
and DR.
%
%f2 #&#
%
\begin{figure*}[b]

\includegraphics{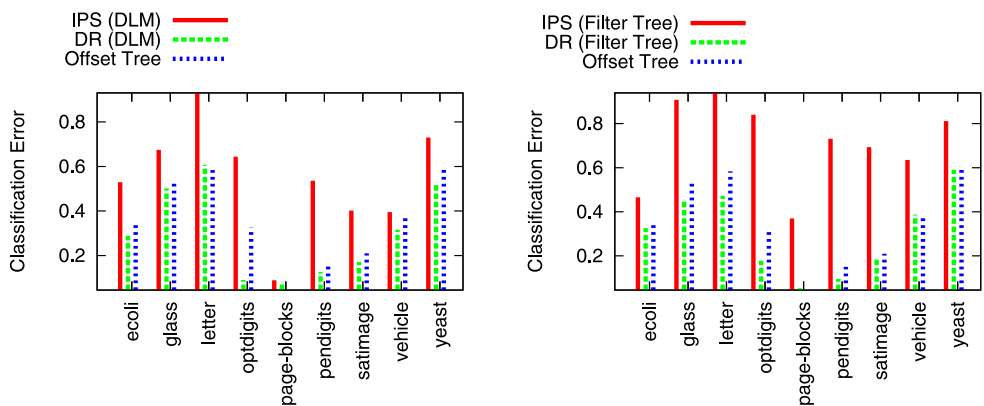}

\caption{Classification error of direct loss minimization (left) and
filter tree (right). Note that the representations used by DLM and
the trees are very different, making any comparison between the two
approaches difficult. However, the Offset Tree and Filter Tree
approaches share a
similar tree representation of the classifiers, so differences in
performance are purely a
matter of superior optimization.}
\label{fig:class-opt}
\end{figure*}

Here, we apply the data transformation in Section~\ref{sec:class-data}
to the \emph{training} data, and then learn a classifier based on the
loss estimated by IPS and DR, respectively. Specifically, for each
dataset, we repeat the following steps $30$ times:
\begin{enumerate}[3.]
\item[1.] We randomly split data into training ($70\%$) and test ($30\%
$) sets;
\item[2.] We apply the transformation in Section~\ref{sec:class-data} to
the training data to obtain a partially labeled set (exploration history);
\item[3.]
We then use the IPS and DR estimators to impute unrevealed losses in
the training data;
that is, we transform each partial-feedback example $(x,a,l)$ into a
\emph{cost sensitive} example
of the form $(x,l_1,\ldots,l_K)$ where $l_{a'}$ is the loss for action
$a'$, imputed from the
partial feedback data as follows:
\[
l_{a'}= %
\cases{ \hat{l}\bigl(x,a'\bigr) +
\displaystyle\frac{l-\hat{l}(x,a')}{\hmu_1(a'\vert x)}, &\mbox{if $a'=a$},
\cr
\hat{l}
\bigl(x,a'\bigr), &\mbox{if $a'\ne a$.} } %
\]
In both cases, $\hmu_1(a'\vert x)=1/K$ (recall that $\hmu_1=\hmu
_k$); in DR we use the loss estimate (described below), in IPS
we use $\hat{l}(x,a')=0$;
\item[4.] Two cost-sensitive multiclass classification algorithms are used
to learn a classifier from the losses completed by either IPS or DR:
the first is DLM used also in the previous section
(see Appendix~\ref{app:dlm} and \citeauthor{McAllester11Direct},
\citeyear{McAllester11Direct}), the
other is the Filter Tree reduction of \citet{Beygelzimer08Multiclass}
applied to a decision-tree base learner (see Appendix~\ref{app:tree}
for more details);
\item[5.] Finally, we evaluate the learned classifiers on the test
data to
obtain classification error. \label{step:class-opt}
\end{enumerate}

Again, we use least-squares ridge regression to build a linear loss
estimator: $\hat{l}(x,a)=w_a\cdot x$. However, since the training data
is partially labeled, $w_a$ is fitted only using training data
$(x,a',l)$ for which $a=a'$. Note that this choice slightly
violates our assumptions, because $\hat{l}$ is \emph{not} independent
of the training data $z_n$.
However, we expect the dependence to be rather weak, and we find this
approach to be more realistic
in practical scenarios where one might want to use all available data
to form the reward estimator, for instance due to data scarcity.

Average classification errors (obtained in Step~5
above) of
$30$ runs are plotted in Figure~\ref{fig:class-opt}. Clearly, for
policy optimization, the advantage of the DR is even greater than for
policy evaluation. In all datasets, DR provides substantially more
reliable loss estimates than IPS, and results in significantly improved
classifiers.

Figure~\ref{fig:class-opt} also includes classification error of the
Offset Tree reduction (\citeauthor{Beygelzimer09Offset}, \citeyear
{Beygelzimer09Offset}), which is designed
specifically for policy
optimization with partially labeled data.\footnote{We used decision
trees as the base learner in Offset Trees to parallel our base learner
choice in Filter Trees. The numbers reported
here are not identical to those by \citet{Beygelzimer09Offset},
even though we used a similar protocol on the same datasets,
probably because of small differences in the data structures used.}
While the IPS versions of DLM and Filter Tree are
rather weak, the DR versions are competitive with Offset Tree in all
datasets, and in some cases significantly outperform Offset Tree.

Our experiments show that DR provides similar improvements in two very
different algorithms, one based on gradient descent, the other based on
tree induction, suggesting the DR technique is generally useful when
combined with different algorithmic choices.

%s4.2 #&#
\subsection{Estimating the Average Number of User Visits}
\label{sec:visit}

The next problem we consider is estimating the average number of user
visits to a popular Internet portal. We formulate this as a regression
problem and in our evaluation introduce an artificial covariate shift. As
in the previous section, the original data is real-world, but the
covariate shift is simulated.
%f3 #&#
%
\begin{figure*}

\includegraphics{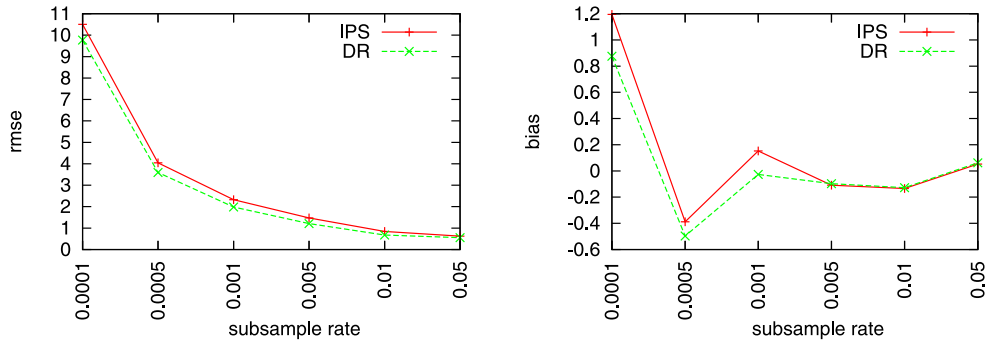}

\caption{Comparison of IPS and DR: \rmse\ (left), \bias\ (right). The
ground truth policy value (average number of user visits) is $23.8$.}
\label{fig:cv}
\end{figure*}

Real user visits to the website
were recorded for about $4$ million \emph{bcookies}\footnote{A
bcookie is a unique string that identifies a user. Strictly speaking,
one user may correspond to multiple bcookies, but for simplicity
we equate a bcookie with a user.} randomly
selected from all bcookies during March 2010. Each bcookie is
associated with a sparse binary covariate vector in $5 000$ dimensions.
These covariates describe browsing behavior as well as other information
(such as age, gender and geographical location) of the bcookie. We
chose a fixed time window in March 2010 and calculated the number of visits
by each selected bcookie during this window. To summarize,
the dataset contains $N=3\mbox{,}854\mbox{,}689$ data points: $D=\{
(b_i,x_i,v_i)\}
_{i=1,\ldots,N}$, where
$b_i$ is the $i$th (unique) bcookie, $x_i$ is the corresponding
binary covariate vector, and $v_i$ is the number of visits (the
response variable); we treat the
empirical distribution over $D$ as the ground truth.

If it is possible to sample $x$ uniformly at random from $D$ and
measure the corresponding value $v$, the sample mean of
$v$ will be an unbiased estimate of the true average number of user
visits, which is $23.8$ in this problem.
However, in various situations, it may be difficult or impossible to
ensure a
uniform sampling scheme due to practical constraints.
Instead, the best that one can do is to sample $x$ from some other
distribution (e.g., allowed by the business constraints)
and measure the corresponding value $v$. In other words, the sampling
distribution of $x$ is changed, but the conditional distribution of $v$
given $x$ remains the same. In this case, the sample average of $v$ may
be a biased estimate of the true quantity of interest.
%We study a setting where we can sample $x\sim D$, but the response $v$
%is provided only for a small fraction of data, and comes from a biased
%distribution $D'(x)$ which differs from $D(x)$.
This setting is known as \emph{covariate shift} (\citeauthor
{Shimodaira00Improving}, \citeyear{Shimodaira00Improving}), where data
are missing at random (see \citeauthor{KangSc07}, \citeyear
{KangSc07}, for related comparisons).

Covariate shift can be modeled as a contextual bandit problem with 2 actions:
action $a=0$ corresponding to ``conceal the response''
and action $a=1$ corresponding to ``reveal the response.'' Below we specify
the stationary exploration policy $\mu_k(a\vert x)=\mu_1(a\vert x)$.
The contextual bandit
data is generated by first sampling $(x,v)\sim D$, then choosing an
action $a\sim\mu_1(\,\cdot\,\vert x)$,
and observing the reward $r=a\cdot v$ (i.e., reward is only revealed if
$a=1$). The exploration policy $\mu_1$ determines the covariate shift.
The quantity of interest, $\E_D[v]$, corresponds to the value of the
constant policy $\nu$ which always chooses ``reveal the response.''

To define the exploration sampling probabilities $\mu_1(a=1\vert x)$,
we adopted an approach similar to \citet{Gretton08Covariate},
with a
bias toward the smaller values along the first principal component of
the distribution over $x$. In particular, we obtained the first
principal component (denoted $\bar{x}$) of all covariate vectors $\{
x_i\}_{i=1,\ldots,N}$, and projected all data onto $\bar{x}$. Let
$\phi$ be the density of a univariate normal distribution with mean
$m+(\bar{m}-m)/3$ and standard deviation $(\bar{m}-m)/4$, where $m$
is the minimum and $\bar{m}$ is the mean of the projected values. We
set $\mu_1(a=1\vert x)=\min\{\phi(x\cdot\bar{x}),1\}$.

%With this large-scale, real-world data, we repeated the following
%steps $100$ times:
%control data size.}
%probability $p_i$, and $a_i=0$ otherwise.}

To control the size of exploration data, we randomly subsampled a
fraction $f\in\{0.0001$, $0.0005$, $0.001$, $0.005$, $0.01$, $0.05\}$
from the entire dataset $D$ and then chose actions $a$ according to the
exploration policy. We then calculated the IPS and DR estimates on this
subsample, assuming perfect logging, that is, $\hmu_k=\mu
_k$.\footnote{Assuming perfect knowledge of exploration probabilities
is fair when we compare IPS and DR. However, it does not give
implications of how DR compares against DM when there is an estimation
error in $\hmu_k$.} The whole process was repeated $100$ times.

The DR estimator required building a reward model $\hr(x,a)$, which,
for a given covariate vector $x$ and $a=1$, predicted the average
number of visits (and for $a=0$ was equal to zero). Again,
least-squares ridge regression was used on a separate dataset to fit a
linear model $\hr(x,1)=w\cdot x$ from the exploration data.
%
% MD: what separate dataset?

%PCA sampling trick as in the previous problem to compute the
%probability of each bcookie being subsampled. On average, the
%subsampling probability is around $1/4$. The set of subsampled
%bcookies is called the ``treatment group''. Treatment groups sampled
%this way are indeed biased: in our dataset, the average number of
%visits in a treatment group is around $29.5$, in contrast to the
%actual number $23.8$, resulting in a huge bias of $24\%$. Propensity
%scoring methods are therefore necessary to eliminate this sampling
%bias.

Figure~\ref{fig:cv} summarizes the estimation error of the two methods
with increasing exploration data size. For both IPS and DR, the
estimation error goes down with more data. In terms of \rmse, the DR
estimator is consistently better than IPS, especially when dataset size
is smaller. The DR estimator often reduces the \rmse\  by a fraction
between $10\%$ and $20\%$, and on average by $13.6\%$. By comparing to
the \bias\ values (which are much smaller), it is clear that DR's gain
of accuracy comes from a lower variance, which accelerates convergence
of the estimator to the true value. These results confirm our analysis
that DR tends to reduce variance provided that a reasonable reward
estimator is available.

%s4.3 #&#
\subsection{Content Slotting in Response to User Queries}

In this section, we compare our estimators
on a proprietary real-world
dataset consisting of web search queries. In response to a search query,
the search engine returns a set of search results. A search result
can be of various types such as a web-link, a news snippet or a movie
information
snippet. We will be evaluating policies that decide which among the
different result types to present at the first position.
The reward is meant to capture the relevance for the user. It equals
$+1$ if the
user clicks on the result at the first position, $-1$ if the user
clicks on some result below the first position,
and $0$ otherwise (for instance, if the user leaves the search page, or decides
to rewrite the query). We call this a \emph{click-skip reward}.

Our partially labeled dataset consists of tuples of the form
$(x_k,a_k,r_k,p_k)$, where $x_k$ is the covariate vector (a~sparse,
high-dimensional representation of the terms of the query as well as
other contextual information, such as user information), $a_k
\in\{$web-link, news, movie$\}$ is the type of result at the first position,
$r_k$ is the click-skip reward, and $p_k$ is
the recorded probability with which the exploration policy chose the
given result type.
Note that due to practical constraints, the values $p_k$ do not always
exactly correspond to
$\mu_k(a_k\vert x_k)$ and should be really viewed as the ``best
effort'' approximation of perfect logging.
We still expect them to be highly accurate, so we use the estimator
$\hmu_k(a_k\vert x_k)=p_k$.

%Note that in general, $\mu(a|x)$ can
%be zero (for instance, when a given back-end is not available or
%returns no
%content), but $\sum_a \mu(a|x)$ is always equal to $1.0$.

The page views corresponding to these tuples represent a small
percentage of user traffic to a major website; any visit to the website
had a
small chance of being part of this experiment. Data was
collected over a span of several days during July 2011. It consists
of 1.2 million tuples, out of which the first 1 million were used
for estimating $\hr$ (training data) with the remainder used for policy
evaluation (evaluation data). The evaluation data was further
split into 10 independent subsets of equal size, which were
used to estimate variance of the compared estimators.\looseness=-1
%For estimating the variance of the compared
%methods, the evaluation data was divided into 10 independent
%subsets of equal size.

We estimated the value of two policies: the exploration policy itself,
and the \emph{argmax} policy (described below).
Evaluating exploration policy on its own exploration data (we call this
setup \emph{self-evaluation}) serves as a sanity check. The \emph
{argmax} policy
is based on a linear estimator $r'(x,a)=w_a\cdot x$ (in general
different from $\hr$),
and chooses the action with the largest predicted reward $r'(x,a)$
(hence the name).
We fitted $r'(x,a)$ on training data by
importance-weighted linear regression with importance weights $1/p_k$.
Note that both $\hr$ and $r'$ are linear estimators
obtained from the same training set, but $\hr$ was computed without
importance weights
and we therefore expect it to be more biased.

%Note also that
%logging the true correct probabilities could be a challenge in
%practice due to various engineering limitations, so instead
%we resorted to estimating them from our data, by building an
%action-specific
%conditional probability estimate. This $\hat{p_k}$ estimate uses the
%original, but
%possibly noisy and biased original value $p_k$ as a feature.

Table~\ref{fig:fedsci_comparison} contains the comparison of
IPS, DM and DR, for both policies under consideration.
For business reasons, we do not report the
estimated reward directly, but normalize to either the empirical
average reward
(for \emph{self-evaluation}) or the IPS estimate (for the
\emph{argmax} policy evaluation).
%
%In both cases, the RS estimate has a much larger variance than the
%other estimators. Note that the minimum observed $p_k = 1/13$,
%which should indicate to us that a naive rejection sampling approach
%would suffer from the data efficiency problem. Indeed, out of
%approx. 20 000 samples per evaluation subset, about 900 are added
%to the history for the argmax policy. In contrast, the DR method
%adds about 13 000 samples, a factor of 14 improvement.
%
%t2 #&#
%
\begin{table}[t]
\tablewidth=235pt
\tabcolsep=0pt
\caption{The results of different policy evaluators on two standard
policies for a real-world exploration problem. In the first column,
results are normalized by the (known) actual reward of the deployed
policy. In the second column, results are normalized by the reward
reported by IPS. All $\pm$ are computed as standard deviations over
results on 10 disjoint test sets. In previous publication of the same
experiments (Dud{\'i}k et~al., \citeyear{DELL12}), we used a
deterministic-policy version of
DR (the same as in Dud{\'i}k, Langford and Li, \citeyear{DLL11}),
hence the results for
self-evaluation presented there slightly differ}
\label{fig:fedsci_comparison}
\begin{tabular*}{235pt}{@{\extracolsep{\fill}}lcc@{}}
\hline
& \textbf{Self-evaluation} & \textbf{Argmax} \\
\hline
IPS & $0.995\pm0.041$ & $1.000\pm0.027$ \\
DM & $1.213\pm0.010$ & $1.211\pm0.002$ \\
DR & $0.974\pm0.039$ & $0.991\pm0.026$ \\
\hline
\end{tabular*}
\end{table}

%Note that in the case of \emph{self-evaluation}, $\pi(a|x) =
%majority of the evaluation methods are substantially simplified (in
%particular,
%Eval now accepts all samples), whereas for \emph{argmax}, $\pi(a|x) =
%1.0$ if $a$ is
%the action with the highest predicted reward and $0$ otherwise (since
%a deterministic policy). Note also that in the case of \emph{argmax}
%evaluation, we
%restrict the action space for a given $x$ to those actions for which
%$\mu(a|x)>0$.

The experimental results are generally in line with theory. The
variance is smallest for DR, although IPS does surprisingly
well on this dataset, presumably because $\hr$ is not sufficiently
accurate. The Direct Method (DM) has an unsurprisingly large bias.
If we divide the listed standard deviations by $\sqrt{10}$, we obtain
standard errors, suggesting that DR has a slight bias (on self-evaluation
where we know the ground truth). We believe that this is due to imperfect
logging.

%In any case, \DRns dominates RS in
%terms of variance as it was designed to do, and has smaller bias and
%variance than DR.

%s5 #&#
\section{Evaluation of Nonstationary Policies} \label{sec:nonstationary}

%s5.1 #&#
\subsection{Problem Definition}

The contextual bandit setting can also be used to model a broad class
of sequential decision-making problems, where the decision maker adapts
her action-selection policy over time, based on her observed history of
context-action-reward triples. In contrast to policies studied in the
previous two sections, such a policy depends on both the current
context and the current \emph{history} and is therefore \emph{nonstationary}.

In the personalized news recommendation example (\citeauthor
{Li10Contextual}, \citeyear{Li10Contextual}), a learning algorithm
chooses an article (an action)
for the current user (the context), with the need for balancing \emph
{exploration} and \emph{exploitation}. Exploration corresponds to
presenting articles about which the algorithm does not yet have enough
data to conclude if they are of interest to a particular type of user.
Exploitation corresponds to presenting articles for which the algorithm
collected enough data to know that they elicit a positive response. At
the beginning, the algorithm may pursue more aggressive exploration
since it has a more limited knowledge of what the users like.
%of which article appears most interesting to a particular type of user.
As more and more data is collected, the algorithm eventually converges
to a good recommendation policy and performs more exploitation.
Obviously, for the same user, the algorithm may choose different
articles in different stages, so the policy is \emph{not} stationary.
In machine learning terminology, such adaptive procedures are called
\emph{online learning} algorithms. Evaluating performance of an online
learning algorithm (in terms of average per-step reward when run for
$T$ steps) is an important problem in practice. Online learning
algorithms are specific instances of nonstationary policies.

%We now shift focus to evaluation of \emph{nonstationary} policies. In
%contrast to stationary ones, a nonstationary policy maps the current
%context \emph{and} a history of past rounds to an action (or a
%distribution of actions if the policy is randomized).
%
%While stationary policies are often sufficient in batch learning,
%in online learning
%with partial feedback, the most successful policies need to remember
%the past; that is,
%they are nonstationary.
%
%The most important example of nonstationary policies is probably
%online-learning algorithms (e.g., \citet{ACF02} and

%
%Nonstationary policies are particularly useful in interactive
%settings, when the policy can automatically adapt to observations,
%and, in partial feedback settings, also automatically optimize the
%what is known to gain immediate reward and exploring what is not
%known to potentially gather higher rewards in the future. Such
%policies fall under the umbrella of \emph{online-learning algorithms}
%which are applied to problems like news article recommendation (e.g.,

Formally, a nonstationary randomized policy is described by a
conditional distribution $\pi(a_t\vert x_t,h_{t-1})$
of choosing an action $a_t$ on a context $x_t$, given the history of
past observations
\[
h_{t-1} = (x_1,a_1,r_1),\ldots
,(x_{t-1},a_{t-1},r_{t-1}) .
\]
We use the index $t$ (instead of $k$), and write $h_t$ (instead
of $z_k$) to make clear the distinction between the histories
experienced by the target
policy $\pi$ versus the exploration policy $\mu$.

A target history of length $T$ is denoted $h_T$. In our analysis, we
extend the
target policy $\pi(a_t\vert x_t,h_{t-1})$
into a probability distribution
over $h_T$ defined by the factoring
\[
\pi(x_t,a_t,r_t\vert h_{t-1}) =
D(x_t)\pi(a_t\vert x_t,h_{t-1})D(r_t
\vert x_t,a_t) .
\]
Similarly to
$\mu$, we define shorthands $\pi_t(x,a,r)$, $\Pr_t^\pi$, $\E_t^\pi$.
The goal of nonstationary policy evaluation is to estimate the expected
cumulative
reward of policy $\pi$ after $T$ rounds:
\[
V_{1:T}=\EE_{h_T\sim\pi} \Biggl[\sum
_{t=1}^T r_t \Biggr] .
\]
In the news recommendation example, $r_t$ indicates whether a user
clicked on the recommended article, and $V_{1:T}$ is the expected
number of clicks garnered by an online learning algorithm after serving
$T$ user visits. A more effective learning algorithm, by definition,
will have a higher $V_{1:T}$ value (\citeauthor{Li10Contextual},
\citeyear{Li10Contextual}).
%measures the average click probability per user visit, one of the
%major metrics with critical business importance.

Again, to have unbiased policy evaluation, we assume
that if $\pi_t(a\vert x)>0$ for any $t$ (and some history $h_{t-1}$)
then $\mu_k(a\vert x)>0$ for all $k$ (and all possible histories
$z_{k-1}$). This clearly holds
for instance if $\mu_k(a\vert x)>0$ for all $a$.

In our analysis of nonstationary policy evaluation, we assume perfect
logging, that is, we assume access to probabilities
\[
p_k\coloneqq\mu_k(a_k\vert
x_k) .
\]
Whereas in general this assumption does not hold, it is realistic in
some applications such as those on the Internet.
For example, when a website chooses one news article from a pool to
recommend to a user, engineers often have full
control/knowledge of how to randomize the article selection process
(\citeauthor{Li10Contextual}, \citeyear{Li10Contextual}; \citeauthor
{LCLW11}, \citeyear{LCLW11}).

%s5.2 #&#
\subsection{Relation to Dynamic Treatment Regimes}
\label{sec:dtr-nonstationary}

The nonstationary policy evaluation problem defined above is closely
related to DTR analysis in a longitudinal observational study. Using
the same notation, the inference goal in DTR is to estimate the
expected sum of rewards by following a possibly randomized rule $\pi$
for $T$ steps.\footnote{In DTR often the goal is to estimate the
expectation of a \emph{composite} outcome that depends on the entire
length-$T$ trajectory. However, the objective of composite outcomes can
easily be reformulated as a sum of properly redefined rewards.}
%
%V_{1:T}=\E_{\pi}\left[\sum_{t=1}^T r_t\right].
%
Unlike contextual bandits, there is no assumption on the distribution
from which the data $z_n$ is generated. More precisely,
%with a little abuse of notation,
given an exploration policy $\mu$, the data generation is described by
\begin{eqnarray*}
&&\mu(x_k,a_k,r_k|z_{k-1})\\
&&\quad=D(x_k|z_{k-1})
\mu(a_k|x_k,z_{k-1})D(r_k|x_k,a_k,z_{k-1})
.
\end{eqnarray*}
Compared to the data-generation process in contextual bandits (see
Section~\ref{sec:definition}),
one allows the laws of $x_k$ and $r_k$ to depend on history $z_{k-1}$.
The target policy $\pi$ is subject to the same conditional laws.
%where $D(\cdot)$ are unknown distributions.
The setting in longitudinal observational studies is therefore
%be described as the set $\{(x_1,a_1,r_1,\ldots,x_T,a_T,r_T)_i\}_{i=1,2,
%
%a Markov decision process with horizon $T$, and is
more general than contextual bandits.
%
% MD: do we want to state this for pi or for mu?

IPS-style estimators (such as DR of the previous section) can be
extended to handle nonstationary policy evaluation, where the
likelihood ratios are now the ratios of likelihoods of the whole
length-$T$ trajectories. In DTR analysis, it is often assumed that the
number of trajectories is much larger than $T$. Under this assumption
and with $T$ small, the variance of IPS-style estimates is on the order
of $O(1/n)$, diminishing to $0$ as $n\rightarrow\infty$.

In contextual bandits, one similarly assumes $n\gg T$. However, the
number of steps $T$ is often large, ranging from hundreds to millions.
The likelihood ratio for a length-$T$ trajectory can be exponential in
$T$, resulting in exponentially large variance. As a concrete example,
consider the case where the exploration policy (i.e., the treatment
mechanism) chooses actions uniformly at random from $K$ possibilities,
and where the target policy $\pi$ is a deterministic function of the
current history and context. The likelihood ratio of any trajectory is
exactly $K^T$, and there are $n/T$ trajectories (by breaking $z_n$ into
$n/T$ pieces of length $T$). Assuming bounded variance of rewards, the
variance of IPS-style estimators given data $z_n$ is $O(TK^{T}/n)$,
which can be extremely large (or even vacuous) for even moderate values
of $T$, such as those in the studies of online learning in the Internet
applications.
%
% except in the practically unlikely situation where $n$ is
%exponentially larger than $T$.

In contrast, the ``replay'' approach of \citet{LCLW11} takes advantage
of the independence between $(x_k,r_k)$ and history $z_{k-1}$. It has a
variance of $O(KT/n)$, ignoring logarithmic terms, when the exploration
policy is uniformly random.
%
%Their variance, however, can be very large, since the importance ratio
%may be exponentially large in $T$ and $T$ is often large in contextual
%bandit problems.
%because
%they do not give a prescription for generating target histories.
%However, if actions have equal probability of being chosen during
%exploration (that is, $\mu_k(a_k|x_k)\equiv1/K$ for all $K$), then the
%``replay'' approach of \citet{LCLW11} may be used to simulate the
%policy $\pi$ and obtain an unbiased estimate of its value with smaller
%variance.
%
When the exploration data is generated by a nonuniformly random policy,
one may apply rejection sampling to simulate uniformly random
exploration, obtaining a subset of the
exploration data, which can then be used to run the replay approach.
%
%rejection sampling can provide unbiased samples of target histories
%that can be used in the same way as \citet{LCLW11},
However, this method may discard a large fraction of data, especially
when the historical actions in the log are chosen
from a highly nonuniform distribution, which can yield
%very short histories and
an unacceptably large variance. The next subsection describes an
improved replay-based estimator that uses doubly-robust estimation as
well as a variant of rejection sampling.

%s5.3 #&#
\subsection{A Nonstationary Policy Evaluator}
\label{sec:evaluator}

\begin{algorithm}[h]
\caption{\newline DR-ns($\pi$, $\{(x_k,a_k,r_k,p_k)\}_{k=1,2,\ldots
,n}$, $\hr
$, $\rho$, $c_{\max}$, $T$)}
\label{algo:eval}
\flushleft
Input:
\begin{enumerate}[]
\item[] target nonstationary policy $\pi$
\item[] exploration data $\{(x_k,a_k,r_k,p_k)\}_{k=1,2,\ldots,n}$
\item[] reward estimator $\hr(x,a)$
\item[] rejection sampling parameters:

\hspace*{3pt}$q\in[0,1]$ and $c_{\max}\in(0,1]$
\item[] number of steps $T$ for estimation
\end{enumerate}
Initialize:
\begin{enumerate}[]
\item[] simulated history of target policy $h_0\gets\emptyset$
\item[] simulated step of target policy $t\gets0$
\item[] acceptance rate multiplier $c_1\gets c_{\max}$
\item[] cumulative reward estimate $\hV_\DRns\gets0$
\item[] cumulative normalizing weight $C\gets0$
\item[] importance weights seen so far $Q\gets\emptyset$
\end{enumerate}
For $k=1,2,\ldots$ consider event $(x_k,a_k,r_k,p_k)$:
\renewcommand\theenumi{(\arabic{enumi})}
\renewcommand\labelenumi{\theenumi}
\begin{enumerate}[(3)]
\item
\label{step:DR:k}
$
\hV_{k}\gets\hat{r}(x_k,\pi_t)
+ \frac{\pi_t(a_k\vert x_k)}{p_k}
\cdot(r_k-\hr(x_k,a_k))
$
\item
\label{step:update}
$\hV_\DRns\gets\hV_\DRns+ c_t \hV_{k}$
\item$
C\gets C+c_t$
\item$
Q\gets Q\cup\{\frac{p_k}{\pi_t(a_k\vert x_k)} \}$\vspace*{3pt}
\item
\label{step:sample}
Let $u_k\sim\operatorname{\textsc{uniform}}[0,1]$\vspace*{2pt}
\item
\label{step:accept}
If $
u_k\le\frac{c_t\pi_t(a_k\vert
x_k)}{p_k}$
\begin{enumerate}[(b)]
\item[(a)]$h_t\gets h_{t-1}+(x_k,a_k,r_k)$
\item[(b)]$t\gets t+1$
\item[(c)] if $t=T+1$, go to ``Exit''
\item[(d)]$c_t\gets\min\{c_{\max}, \rho\mbox{th quantile of } Q\}$
\label{step:rho}
\end{enumerate}
\end{enumerate}
Exit: If $t<T+1$, report failure and terminate; otherwise, return:
\begin{enumerate}[]
\item[] cumulative reward estimate $\hV_\DRns$
\item[] average reward estimate $\hat{V}^{\mathrm{avg}}_\DRns
\coloneqq\hV_\DRns/C$
\end{enumerate}
\end{algorithm}

Our replay-based nonstationary policy evaluator (Algorithm \ref
{algo:eval}) takes advantage
of
high accuracy of DR estimator while tackling nonstationarity via
rejection sampling.
We substatially improve sample use (i.e., acceptance rate) in rejection sampling
while only modestly increasing the bias.
This algorithm is
referred to as DR-ns, for ``doubly robust nonstationary.''
Over the
run of the algorithm, we process the exploration history
and run rejection sampling [Steps \ref{step:sample}--\ref
{step:accept}] to create a simulated
history $h_t$ of the interaction between the target policy
and the environment. If the algorithm manages to simulate $T$ steps of
history, it exits and returns an estimate
$\hV_\DRns$ of the cumulative reward $V_{1:T}$,
and an estimate $\hat{V}^{\mathrm{avg}}_\DRns$ of the average reward
$V_{1:T}/T$;
otherwise, it reports failure indicating not enough data is available.

Since we assume $n\gg T$, the algorithm fails with a small probability
as long as the exploration policy does not assign too small
probabilities to actions. Specifically, let $\alpha>0$ be a lower
bound on the acceptance probability in the rejection sampling step;
that is, the condition in Step \ref{step:accept} succeeds with
probability at least $\alpha$. Then, using the Hoeffding's inequality,
one can show that
the probability of failure of the algorithm is at most $\delta$ if
\[
n \ge\frac{T+\ln(e/\delta)}{\alpha} . %\frac{2}{\alpha}\left(T+\ln
\]

Note that the algorithm returns one ``sample'' of the policy value. In
reality, the algorithm continuously consumes a stream of $n$ data,
outputs a sample of policy value whenever a length-$T$ history is
simulated, and finally returns the average of these samples. Suppose we
aim to simulate $m$ histories of length $T$. Again, by Hoeffding's
inequality, the probability of failing to obtain $m$ trajectories is at
most $\delta$ if
\[
n \ge\frac{mT+\ln(e/\delta)}{\alpha} . %\frac{2}{\alpha}\left(mT+
\]

Compared with naive rejection sampling, our approach differs in two respects.
First, we use not only the accepted samples, but also the rejected ones
to estimate the expected reward $\E^\pi_t[r]$ with a DR estimator
[see Step~\ref{step:DR:k}]. As we will see below, the value of $1/c_t$
is in expectation equal to the total number of exploration samples used
while simulating the $t$th action of the target policy. Therefore, in
Step~\ref{step:update}, we effectively take an average of $1/c_t$
estimates of $\E^\pi_t[r]$,
decreasing the variance of the final estimator. This is in addition to
lower variance
due to the use of the doubly robust estimate in Step~\ref{step:DR:k}.

The second modification is in the control of the acceptance rate (i.e.,
the bound $\alpha$ above).
When simulating the $t$th action of the target policy,
we accept exploration samples with a probability $\min\{1,c_t\pi
_t/p_k\}$ where
$c_t$ is a multiplier [see Steps \ref{step:sample}--\ref
{step:accept}]. We will see below that the bias of the estimator is
controlled by the probability
that $c_t\pi_t/p_k$ exceeds 1, or equivalently, that $p_k/\pi_t$
falls below $c_t$. As a heuristic toward controlling this probability,
we maintain a set $Q$ consisting
of observed density ratios $p_k/\pi_t$, and at the beginning of
simulating the $t$th action, we set $c_t$ to the $\rho$th quantile of
$Q$, for some small value of $\rho$ [Step~\ref{step:rho}(d)], while
never allowing it to exceed some predetermined $c_{\max}$. Thus, the
value $\rho$
approximately corresponds to the probability value that we wish to control.
Setting $\rho=0$, we obtain the unbiased case (in the limit). By using
larger values of $\rho$, we increase the bias, but reach the length
$T$ with fewer exploration samples thanks to increased acceptance rate.
A similar effect is obtained by varying $c_{\max}$, but the control is
cruder, since it ignores the evaluated policy.
In our experiments, we therefore set $c_{\max}=1$ and rely on $\rho$
to control the acceptance rate. It is an interesting open question how
to select $\rho$ and $c$ in practice.

To study our algorithm DR-ns, we modify the definition of the
exploration history
so as to include the samples $u_k$ from the uniform distribution used
by the algorithm when processing the $k$th exploration sample.
Thus, we have an augmented definition
\[
z_k=(x_1,a_1,r_1,u_1,
\ldots,x_k,a_k,r_k,u_k) .
\]
With this in mind, expressions $\Pr^\mu_k$ and $\E^\mu_k$ include
conditioning on variables $u_1,\ldots,u_{k-1}$, and $\mu$ is viewed
as a
distribution over augmented histories~$z_n$.

For convenience of analysis, we assume in this section that we have
access to an infinite exploration history $z$ (i.e., $z_n$ for
$n=\infty$) and that the counter $t$ in the pseudocode eventually
becomes $T+1$ with probability one (at which point $h_T$ is generated).
Such an assumption is mild in practice when $n$ is much larger than $T$.
%a continuous running
%of our algorithm on an infinite history $z$.

Formally, for $t\ge1$, let $\kappa(t)$ be the index of the $t$th
sample accepted in
Step~\ref{step:accept}; thus, $\kappa$ converts an index in the target
history into an index in the exploration history. We set $\kappa(0)=0$ and
define $\kappa(t)=\infty$ if fewer than $t$ samples are accepted. Note
that $\kappa$ is a deterministic function of the history $z$ (thanks to
including samples $u_k$ in $z$).
We assume that $\Pr_\mu[\kappa(T)=\infty]=0$.
%for every $t$, $\Pr_\mu[\kappa(t)=\infty]=0$.
This means that the algorithm (together with the exploration policy
$\mu$)
generates a distribution over histories $h_T$;
we denote this distribution $\hpi$.

Let $B(t)=\{\kappa(t-1)+1,\kappa(t-1)+2,\ldots,\kappa(t)\}$ for
$t\ge1$ denote the set of sample indices between the $(t-1)$st
acceptance and the $t$th acceptance. This set of
samples is called the $t$th block.
%The inverse operator identifying the block of the $k$-th sample is
% \tau(k) = t\mbox{such that }k\in B(t)
%.
%
The contribution of the $t$th block to the value estimator is denoted
$\hV_{B(t)}=\sum_{k\in B(t)} \hV_{k}$. After completion
of $T$ blocks, the two estimators returned by our algorithm are
\[
\hV_\DRns=\sum_{t=1}^T
c_t \hV_{B(t)} ,\quad \hat{V}^{\mathrm{avg}}_\DRns=
\frac{\sum_{t=1}^T c_t \hV
_{B(t)}}{\sum_{t=1}^T
c_t\vert B(t)\vert} .
\]

%s5.4 #&#
\subsection{Bias Analysis}

%Our goal is to develop an accurate estimator.
%Ideally, we would like to bound the error as a function of an
%increasing number of exploration samples.
%For a nonstationary policy, it can be easily shown that a single
%simulation history of
%that policy can yield a per-step reward that is bounded away by 0.5
%from the expected per-step reward
%regardless of the length of simulation (see, e.g., Example~3 of
%Hence, even for unbiased methods, we cannot accurately estimate the
%expected reward from a single trace.

A simple approach to evaluating a nonstationary policy is to divide the
exploration data
into several parts, run the algorithm separately on each
part to
generate simulated histories, obtaining estimates
$\hV_\DRns^{(1)}, \ldots, \hV_\DRns^{(m)}$, and return the
average $\sum_{i=1}^m \hV_\DRns^{(i)}/m$.\footnote{We only consider
estimators for cumulative rewards (not average rewards) in this
section. We assume that the division into
parts is done sequentially, so that individual estimates are built from
nonoverlapping sequences of $T$ consecutive blocks of examples.}
Here, we assume $n$ is large enough so that $m$ simulated histories of
length $T$ can be generated with high probability.
Using
standard concentration inequalities, we can then show that the average is
within $O(1/\sqrt{m})$ of the expectation $\E_\mu[\hV_\DRns]$.
The remaining piece is then bounding the bias term $\E_\mu[\hV_\DRns
]-\E_\pi[\sum_{t=1}^T r_t ]$.\footnote{As shown in \citet
{LCLW11}, when $m$ is constant, making $T$
large does not necessarily reduce variance of any estimator of
nonstationary policies.}
%Interested readers are referred to Appendix~\ref{app:PE} for a
%technique that extracts a stationary policy from a nonstationary
%policy.
%The variance of this stationary policy's value estimate can decrease
%at a fast rate as $T\rightarrow\infty$.}

%It is easy to prove a lower bound (see section \ref{sec:dev}) showing
%that no large-deviation result holds for the output of Eval when the
%input is a nonstationary policy. This implies that the best we can
%prove is unbiasedness, as was done for the rejection sampling
%evaluator \cite{LCLS10}. In general, perfect unbiasedness is difficult
%to achieve while still taking full advantage of the data, thus we
%prove a theorem here showing how the bias due to a misestimate of $c$
%affects the final output. In this section, we derive a theorem
%bounding how much bias is introduced in the worst case.

Recall that
$\hV_\DRns=\sum_{t=1}^T c_t \hV_{B(t)}$.
The source of bias are events when
$c_t$ is not small enough to guarantee that $c_t\pi_t(a_k\vert x_k)/p_k$
is a probability.
In this case, the probability that the $k$th exploration sample includes
the action $a_k$ and is accepted is
%
%e5.1 #&#
%
\begin{equation}
\label{eq:accepted} p_k \min\biggl\{1, \frac{c_t\pi
_t(a_k|x_k)}{p_k} \biggr\} =
\min\bigl\{p_k, c_t\pi_t(a_k|x_k)
\bigr\},\hspace*{-27pt}
\end{equation}
which may violate the unbiasedness requirement of rejection sampling,
requiring that the probability of acceptance be proportional to
$\pi_t(a_k\vert x_k)$.

Conditioned on $z_{k-1}$ and the induced target history $h_{t-1}$, define
the event
\[
\event_k \coloneqq\bigl\{(x,a)\st c_{t}
\pi_{t}(a\vert x) > \mu_{k}(a\vert x) \bigr\} ,
\]
which contributes to the bias of the estimate, because it corresponds
to cases when the minimum in equation (\ref{eq:accepted}) is attained
by $p_k$.
Associated with this event is the ``bias mass'' $\eps_k$, which
measures (up to scaling by $c_t$) the
difference between the probability of the bad event under $\pi_t$ and
under the run of our algorithm:
\[
\eps_k\coloneqq\PP_{(x,a)\sim\pi_{t}}[\event_k] -
\PP_{(x,a)\sim
\mu_{k}}[\event_k]/c_{t} .
\]
Notice that from the definition of $\event_k$, this mass is
nonnegative. Since the first term is a probability, this mass is at most
$1$. We will assume that this mass is bounded away from $1$, that is,
that there exists $\eps$ such that
for all $k$ and $z_{k-1}$
\[
0\le\eps_k\le\eps<1 .
\]
The following theorem analyzes how much
bias is introduced in the worst case, as a function of $\eps$.
It shows how the bias mass controls the bias of our estimator.
%Its purpose is to identify the key quantities that contribute to the
%bias (the bias mass),
%and to provide insights of what to optimize in practice.
%
%th5.1 #&#
%
\begin{theorem}
\label{thm:main}
For $T\ge1$,
\begin{eqnarray*}
&&\Biggl\vert\E_\mu\Biggl[\sum_{t=1}^T
c_t \hV_{B(t)} \Biggr] -\E_\pi\Biggl[\sum
_{t=1}^T r_t \Biggr]\Biggr\vert\\
&&\quad\le
\frac{T(T+1)}{2} \cdot\frac{\eps}{1-\eps} .
\end{eqnarray*}
\end{theorem}

Intuitively, this theorem says that if a bias of $\eps$ is
introduced in round $t$, its effect on the sum of rewards can be felt
for $T-t$ rounds. Summing over rounds, we expect to get an $O(\eps
T^2)$ effect on the estimator of the cumulative reward.
%or equivalently a bias of $O(\eps T)$ on the average reward.
In
general a very slight bias can
result in a significantly better acceptance rate, and hence
more replicates $\hV_\DRns^{(i)}$.

This theorem is the first of this sort for policy evaluators, although
the mechanics of its proof have appeared in
model-based reinforcement-learning (e.g., \citeauthor{E3}, \citeyear{E3}).
%in Markov decision processes
%A key difference
%here is that we depend on a context with unbounded complexity rather
%than a finite state space.

To prove the main theorem, we state
two technical lemmas bounding the differences of probabilities
and expectations under the target policy and our algorithm
(for proofs of lemmas, see Appendix~\ref{app:proofs2}). The theorem follows
as their immediate consequence. Recall
that $\hpi$ denotes the distribution over target histories generated
by our algorithm (together with the exploration policy $\mu$).
%
%le5.2 #&#
%
\begin{lemma}
\label{LEMMA:COND}
Let $t\le T$, $k\ge1$ and let $z_{k-1}$ be such that the $k$th
exploration sample marks the beginning of the $t$th block,
that is, $\kappa(t-1)=k-1$.
Let $h_{t-1}$ and $c_{t}$ be the target history and
acceptance rate multiplier induced by $z_{k-1}$.
Then:
\begin{eqnarray*}
\sum_{x,a}\bigl\llvert\Pr^\mu_k[x_{\kappa(t)}=x,a_{\kappa(t)}=a]
- \pi_t(x,a) \bigr\rrvert&\le&\frac{2\eps}{1-\eps} ,
\\
\bigl\llvert c_t\E^\mu_k[
\hV_{B(t)}] - \E^\pi_t[r]\bigr\rrvert&\le&
\frac{\eps}{1-\eps} .
\end{eqnarray*}
\end{lemma}
%
%le5.3 #&#
%
\begin{lemma}
\label{LEMMA:MARGINAL}
\[
\sum_{h_T} \bigl\vert\hpi(h_T) - \pi(h_T)\bigr\vert
\le
(2\eps T) / (1-\eps)
.
\]
\end{lemma}
%
%We are now ready to prove Theorem~\ref{thm:main}, bounding the bias of
%our estimator.
%
\begin{pf*}{Proof of Theorem~\ref{thm:main}}
First, bound
$\vert\E_\mu[c_t\cdot\allowbreak \hV_{B(t)}]-\E_\pi[r_t]\vert$
using the
previous two lemmas,
the triangle inequality and H\"older's inequality:
\begin{eqnarray*}
&&\bigl\llvert\E_\mu[c_t\hV_{B(t)}]-
\E_{\pi}[r_t]\bigr\rrvert\\
&&\quad= \bigl\llvert\E_\mu
\bigl[c_t \E^\mu_{\kappa(t)}[\hV_{B(t)}] \bigr]
- \E_\pi[r_t]\bigr\rrvert
\\
&&\quad\le\bigl\vert\E_\mu\bigl[\E^\pi_t[r_t]
\bigr] - \E_\pi\bigl[\E^\pi_t[r_t]
\bigr] \bigr\vert+ \frac{\eps}{1-\eps}
\\
&&\quad= \biggl\llvert\EE_{h_{t-1}\sim\hpi} \biggl[\E^\pi_t
\biggl[r-\frac{1}2 \biggr] \biggr] - \EE_{h_{t-1}\sim\pi} \biggl[
\E^\pi_t \biggl[r-\frac{1}2 \biggr] \biggr] \biggr
\rrvert\\
&&\hphantom{\quad= \biggl\llvert}{} + \frac{\eps}{1-\eps}
\\
&&\quad\le\frac{1}2\sum_{h_{t-1}} \bigl\vert
\hpi(h_{t-1})-\pi(h_{t-1}) \bigr\vert+ \frac{\eps}{1-\eps}
\\
&&\quad\le\frac{1}2\cdot\frac{2\eps(t-1)}{1-\eps} + \frac{\eps
}{1-\eps} =
\frac{\eps t}{1-\eps} .
\end{eqnarray*}
The theorem now follows by summing over $t$ and using the triangle inequality.
\end{pf*}

\section{Experiments: The Nonstationary Case}
\label{sec:experiment2}

We now study how $\DRns$ may achieve greater sample efficiency than
rejection sampling
through the use of a controlled bias. We
evaluate our estimator on the problem of a multiclass multi-label
classification with partial feedback
using the publicly available dataset \rcv\ (\citeauthor{Lewis04Rcv1},
\citeyear{Lewis04Rcv1}). In this
data, the goal is to predict whether a news article is in one of many
Reuters categories given the contents of the article.
This dataset is chosen instead of the UCI benchmarks in Section~\ref
{sec:experiment1}
because of its bigger size, which is helpful for simulating online
learning (i.e., adaptive policies).

%s6.1 #&#
\subsection{Data Generation}
\label{sec:data-generation}

For multi-label dataset like \rcv, an example has the form $(\tx,Y)$, where
$\tx$ is the covariate vector and $Y\subseteq\{1,\ldots,K\}$
is the \emph{set} of correct class labels.\footnote{%
The reason why we call the covariate vector $\tx$ rather than $x$ becomes
in the sequel.}
In our modeling, we assume that any $y\in Y$ is the correct prediction
for $\tx$.
Similar to Section~\ref{sec:class}, an example
$(\tx,Y)$ may be interpreted as
a bandit event with context $\tx$ and loss $l(Y,a)\coloneqq I(a\notin
Y)$, for every action $a\in\{1,\ldots,K\}$.
A classifier can be interpreted as a \emph{stationary} policy
whose expected loss is its classification error.
In this section, we again aim at evaluating expected policy loss, which
can be understood as negative reward.
For our experiments, we only use the $K=4$ top-level classes
in \rcv, namely $\{C,E,G,M\}$. We take a random selection of $40\mbox{,}000$
data points
from the whole dataset and call the resulting dataset $D$.

To construct a partially
labeled exploration data\-set, we simulate a stationary but nonuniform
exploration policy with a bias toward correct answers. This is
meant to emulate the typical setting where a baseline system already
has a good understanding of which actions are likely best. For each
example $(\tx,Y)$,
a uniformly random value $s(a)\in[0.1,1]$ is assigned independently to
each action $a$,
and the final probability of action $a$ is determined by
\[
\mu_1(a\vert\tx,Y,s) = \frac{0.3 \times s(a)}{\sum_{a'} s(a')} +
\frac{0.7 \times I(a\in Y)}{|Y|} .
\]
Note that this policy
will assign a nonzero probability to every action.
Formally, our exploration policy is a function of an extended context
$x=(\tx,Y,s)$, and
our data generating distribution $D(x)$ includes the generation of the
correct answers $Y$ and values $s$. Of course, we will be evaluating
policies $\pi$ that only get to see $\tx$, but have no access to $Y$
and $s$. Also, the estimator $\hl$ (recall that we are evaluating loss
here, not reward) is purely a function of $\tx$ and $a$. We stress
that in a real-world setting, the exploration policy would not have
access to all correct answers $Y$.
%and our simulation represents the setup when the exploration process
%selects high-reward actions more often.

%s6.2 #&#
\subsection{Evaluation of a Nonstationary Policy}

As described before, a fixed (nonadaptive) classifier can be
interpreted as a stationary policy. Similarly, a~classifier that adapts
as more data arrive is equivalent to a nonstationary policy.

In our experiments, we evaluate performance of an adaptive \emph
{$\epsilon$-greedy} classifier defined as follows: with probability
$\epsilon=0.1$, it predicts a label drawn uniformly at random from $\{
1,2,\ldots,K\}$; with probability $1-\epsilon$, it predicts the best
label according to a linear score (the ``greedy'' label):
\[
\mathop{\argmax}_a \bigl\{ w^t_a\cdot\tx\bigr\} ,
\]
%
%probability $1-\epsilon$} \\ Unif\{1,2,\ldots,K\} & \mbox{with
%probability $\epsilon$} \end{cases},
where $\{w^t_a\}_{a\in\{1,\ldots,K\}}$ is a set of $K$ weight vectors
at time $t$. This design mimics a commonly used $\epsilon$-greedy
exploration strategy for contextual bandits (e.g., \citeauthor
{Li10Contextual}, \citeyear{Li10Contextual}). Weight vectors $w^t_a$
are obtained by fitting a
logistic regression
model for the binary classification problem $a\in Y$ (positive) \vs\
$a\notin Y$ (negative). The data used to fit $w^t_a$ is described
below. Thus,
the greedy label is the most likely label according to the current set
of logistic regression models.
The loss estimator $\hl(\tx,a)$ is also obtained by fitting a
logistic regression model for $a\in Y$ \vs\ $a\notin Y$, potentially
on a different dataset.

We partition the whole data $D$ randomly into three disjoint subsets:
$D_{\init}$ (initialization set), $D_{\valid}$ (validation set), and
$D_{\eval}$ (evaluation set), consisting of $1\%$, $19\%$, and $80\%$ of
$D$, respectively. Our goal in this experiment is to estimate the
expected loss, $V_{1:T}$, of an adaptive policy $\pi$ after $T=300$ rounds.

The full-feedback set $D_\init$ is used to fit the loss estimator $\hl$.

Since $D_\valid$ is a random subset of $D$, it may be used to simulate
the behavior of policy $\pi$ to obtain an \textit{unbiased} estimate
of $V_{1:T}$. We do this by taking an average of $2 000$ simulations of
$\pi$ on random shuffles of the set $D_\valid$. This estimate,
denoted $\bar{V}_{1:T}$, is a highly accurate approximation to (the
unknown) $V_{1:T}$, and serves as our ground truth.

To assess different policy-value estimators, we randomly permute
$D_\eval$ and transform it into a partially labeled set as described
in Section~\ref{sec:data-generation}. On the resulting partially
labeled data, we then evaluate the policy $\pi$ up to round $T$,
obtaining an estimate of $V_{1:T}$. If the exploration history is not
exhausted, we start the evaluation of $\pi$ again, continuing with the
next exploration sample, but restarting from empty target history (for
$T$ rounds), and repeat until we use up all the exploration data. The
final estimate is the average across thus obtained replicates. We
repeat this process (permutation of $D_\eval$, generation of
exploration history, and policy evaluation until using up all
exploration data) 50 times,
so that we can compare the $50$ estimates against the ground truth
$\bar{V}_{1:T}$ to compute bias and standard deviation of a policy-value
estimator.

Finally, we describe in more detail the $\eps$-greedy adaptive
classifier $\pi$ being evaluated:
\begin{itemize}
\item First, the policy is initialized by fitting weights $w_a^0$ on
the full-feedback set $D_\init$ (similarly to $\hl$). This step
mimics the practical situation where one usually has prior information
(in the form of either domain knowledge or historical data) to
initialize a policy, instead of starting from scratch.
\item After this ``warm-start'' step, the ``online'' phase begins: in
each round, the policy observes a randomly selected $\tx$, predicts a
label in an $\epsilon$-greedy fashion (as described above), and then
observes the corresponding $0/1$ prediction loss. The policy is updated
every $15$ rounds. On those rounds, we retrain weights $w_a^t$ for each
action $a$, using the full feedback set $D_\init$ as well as all the
data from the online phase where the policy chose action $a$. The
online phase terminates after $T=300$ rounds.
\end{itemize}

%Finally, to estimate $V_{1:T}$ from exploration data, we turn $D_\eval
%$ into a partially labeled set (as described in Section
%Algorithm \ref{algo:eval} are compared. This process is repeated $50$
%times, so that we can compare the $50$ estimates against the ground
%truth $\bar{V}_{1:T}$ to compute bias and standard deviation of an
%evaluator.

% (the one which we are interested in evaluating). We update policy $
%observed. On policy update, we simply use an enlarged training set
%containing the initial $400$ fully labeled and additional partially
%labeled examples. The offline training set was fixed in all trials.
%The remaining data was split into two portions, the first containing a
%random $80\%$ of $D$ for evaluation, the second containing $19\%$ of
%$D$ to determine the ground truth. The evaluation set was randomly
%permuted and then transformed into a partially labeled set $D'$ on
%which evaluators were compared. The generation of $D'$ was repeated in
%$50$ trials, from which bias and standard deviation of each evaluator
%were obtained. To estimate the ground-truth value of $\pi$, we
%simulated $\pi$ on the randomly shuffled (fully labeled) $19\%$ of
%ground-truth data $2000$ times to compute its average online loss.

%s6.3 #&#
\subsection{Compared Evaluators}
We compared the following evaluators described earlier: DM for direct
method, RS for the unbiased evaluator based on rejection sampling and
``replay'' (\citeauthor{LCLW11}, \citeyear{LCLW11}), and $\DRns$ as
in Algorithm \ref{algo:eval}
(with $c_{\max}=1$). We also tested a variant of $\DRns$, which does
not monitor the quantile, but instead uses $c_t$ equal to $\min_D \mu
_1(a|x)$; we call it $\WC$ since it uses the worst-case (most
conservative) value of $c_t$ that ensures unbiasedness of rejection sampling.

%s6.4 #&#
\subsection{Results}
Table~\ref{tbl:rcv1-adaptive} summarizes the accuracy of different
evaluators in terms of \rmse\ (root mean squared error), \bias\ (the
absolute difference between the average estimate and the ground truth)
and \std\ (standard deviation of the estimates across different runs).
It should be noted that, given the relatively small number of trials,
the measurement of \bias\ is not statistically significant.
%So for instance, it cannot be inferred in a statistically significant
%way from Table~\ref{tbl:rcv1-static} that WC enjoys a lower bias than
%RS.
However, the table provides $95\%$ confidence interval for the \rmse\
metric that allows a meaningful comparison.

%
%t3 #&#
%
\begin{table}[b]
\tablewidth=240pt
\tabcolsep=0pt
\caption{Nonstationary policy evaluation results} \label{tbl:rcv1-adaptive}
\begin{tabular*}{240pt}{@{\extracolsep{\fill}}lccc@{}}
\hline
\textbf{Evaluator} & \textbf{\rmse}\ ($\mathbf{\pm95\%}$ \textbf
{C.I.}) & \textbf{\bias}& \textbf{\std}\\
\hline
DM & $0.0329 \pm0.0007$ & $0.0328$ & $0.0027$ \\
RS & $0.0179 \pm0.0050$ & $0.0007$ & $0.0181$ \\
$\WC$& $0.0156 \pm0.0037$ & $0.0086$ & $0.0132$ \\
$\DRns\ (\rho=0)$ & $0.0129 \pm0.0034$ & $0.0046$ & $0.0122$
\\
$\DRns\ (\rho=0.01)$ & $\mathbf{0.0089} \pm0.0017$ & $0.0065$ &
$0.0062$ \\
$\DRns\ (\rho=0.05)$ & $0.0123 \pm0.0017$ & ${0.0107}$ & ${0.0061}$
\\
$\DRns\ (\rho=0.1)$ & $0.0946 \pm0.0015$ & $0.0946$ & $0.0053$
\\
\hline
\end{tabular*}
\end{table}

It is clear that although rejection sampling is guaranteed to be
unbiased, its variance is usually the dominating part of its \rmse. At
the other extreme is the direct method, which has the smallest variance
but often suffers large bias. In contrast, our method $\DRns$ is able to
find a good balance between the two extremes and, with proper selection
of the parameter $\rho$, is able to make the evaluation results much
more accurate than others.

It is also clear that the main benefit of $\DRns$ is its low variance,
which stems from the adaptive choice of $c_t$ values. By slightly
violating the unbiasedness guarantee, it increases the effective data
size significantly, hence reducing the variance of its evaluation. For
$\rho>0$, $\DRns$ was able to extract many more trajectories of length
$300$ for evaluating $\pi$, while RS and $\WC$ were able to find only
one such trajectory out of the evaluation set. In fact, if we increase
the trajectory length of $\pi$ from 300 to 500, both RS and $\WC$ are
not able to construct a complete trajectory of length 500 and fail the
task completely.
%
%Finally, we examine the use of a reward estimator in Eval. As
%indicated by the large bias of the direct method, the reward estimator
%is not very accurate. However, as is consistent with previous work
%the Eval variant that set $\hr\equiv0$. Eval thus benefits further
%from the doubly robust technique to reduce its evaluation variance.
%
% Old results: in AISTATS submission.
%Evaluator & \rmse& \bias& \std\tabularnewline
%DM & $0.0143$ & $0.0143$ & ${0.0000}$ \tabularnewline\hline
%RS & $0.0138$ & ${0.0014}$ & $0.0137$ \tabularnewline\hline
%WC & $0.0157$ & $0.0051$ & $0.0149$ \tabularnewline\hline
%Eval($\rho=0$) & $0.0038$ & $0.0024$ & $0.0029$ \tabularnewline\hline
%Eval($\rho=0.01$) & $\mathbf{0.0024}$ & $0.0022$ & $0.0010$
%Eval($\rho=0.03$) & $\mathbf{0.0024}$ & $0.0021$ & $0.0012$
%Eval($\rho=0$,$\hr\equiv0$) & $0.0048$ & $0.0022$ & $0.0043$
%
%Old results: AISTATS submission.
%Evaluator & \rmse& \bias& \std\tabularnewline
%DM & $$ & $$ & $$ \tabularnewline\hline
%RS & $0.0179$ & $0.0017$ & $0.0178$ \tabularnewline\hline
%WC & $0.0232$ & $0.0056$ & $0.0225$ \tabularnewline\hline
%Eval($\rho=0$) & $0.0058$ & $0.0033$ & $0.0048$ \tabularnewline\hline
%Eval($\rho=0.001$) & $0.0048$ & $0.0023$ & $0.0042$ \tabularnewline
%Eval($\rho=0.003$) & $\mathbf{0.0035}$ & ${0.0009}$ & ${0.0034}$
%Eval($\rho=0$,$\hr\equiv0$) & $0.0109$ & $0.0024$ & $0.0106$
%
% New results: UAI submission.

%s7 #&#
\section{Conclusions}

Doubly robust policy estimation is an effective technique which
virtually always improves on the widely used inverse propensity score
method. Our analysis shows that doubly robust methods tend to give
more reliable and accurate estimates, for evaluating both stationary
and nonstationary policies. The theory is corroborated by
experiments on benchmark data as well as two large-scale real-world
problems.
In the future, we expect the DR technique to become common
practice in improving contextual bandit algorithms.

% MD: removing the remainder since it seems to speak to the COLT
% audience, and this is not a COLT paper
\iffalse
As an
example, it is interesting to develop a variant of Offset Tree
that can take advantage of better reward models, rather than a
crude, constant reward estimate (\citeauthor{Beygelzimer09Offset},
\citeyear{Beygelzimer09Offset}).

There are definitely opportunities for further improvement. For
example, consider nonstationary policies which can devolve into
round-robin action choices when the rewards are constant (such as the
UCB1 algorithm of \citeauthor{ACF02}, \citeyear{ACF02}). A policy
which cycles through
actions has an expected reward equivalent to a randomized policy which
picks actions uniformly at random. And yet, for such a policy, our
policy evaluator will only accept on average a fraction of $1/K$
uniform random exploration events. How can we build a more
data-efficient policy evaluator for these kinds of situations?
\fi
%[[[TODO: merge in conclusions from the nonstationary paper.]]]

%sA #&#
%
\begin{appendix}

%sA #&#
\section{\texorpdfstring{Proofs of Lemmas \protect\ref{LEMMA:RANGE}--\protect\ref{LEMMA:VAR}}{Proofs of Lemmas 3.1--3.3}}
\label{app:proofs1}

Throughout proofs in this appendix, we write $\hr$ and $r^*$ instead
of $\hr(x,a)$ and $r^*(x,a)$ when $x$ and $a$ are clear from the context,
and similarly for $\Delta$ and $\q_k$.
\renewcommand{\thelemmas}{3.1}{
%leA.1 #&#
%
\begin{lemmas}
The range of $\hV_{k}$ is bounded as
\[
\vert\hV_{k} \vert\le1 + M .
\]
\end{lemmas}
}
\begin{pf}
\begin{eqnarray*}
\vert\hV_{k} \vert&=& \biggl\llvert\hat{r}(x_k,\nu) +
\frac{\nu(a_k\vert x_k)} {\hmu
_k(a_k\vert x_k)} \cdot\bigl(r_k-\hr(x_k,a_k)
\bigr) \biggr\rrvert
\\
&\le& \bigl\llvert\hat{r}(x_k,\nu)\bigr\rrvert+ \frac{\nu
(a_k\vert x_k)}{
\hmu_k(a_k\vert x_k)}
\cdot\bigl\vert r_k-\hr(x_k,a_k) \bigr\vert
\\
&\le& 1 + M,
\end{eqnarray*}
where the last inequality follows because
$\hr$ and $r_k$ are bounded in $[0,1]$.
\end{pf}
\renewcommand{\thelemmas}{3.2}{
%leA.2 #&#
%
\begin{lemmas}
The expectation of the term $\hV_{k}$ is
\[
\E^\mu_k[\hV_{k}] = \EE_{(x,a)\sim\nu}
\bigl[r^*(x,a) + \bigl(1-\q_k(x,a) \bigr)\Delta(x,a) \bigr] .
\]
\end{lemmas}
}
\begin{pf}
%
%eA.1 #&#
%
\begin{eqnarray}
\E^\mu_k [\hV_{k} ] &=&
\EE_{(x,a,r)\sim\mu_k} \biggl[\hat{r}(x,\nu) + \frac{\nu(a\vert
x)}{\mu_k(a\vert x)}\cdot
\q_k\cdot(r-\hr) \biggr]
\nonumber\\
&
=&\EE_{x\sim D} \bigl[\hat{r}(x,\nu) \bigr]\nonumber\\
&&{} +\EE_{x\sim D}
\biggl[\sum_{a\in\A}\mu_k(a\vert x)
\EE_{r\sim
D(\,\cdot\,\vert x,a)} \biggl[\frac{\nu(a\vert x)}{\mu_k(a\vert
x)}\nonumber\\
&&\hspace*{120pt}{}\cdot\q_k\cdot(r-\hr)
\biggr] \biggr]
\nonumber
\\
&
=&\EE_{x\sim D} \bigl[\hat{r}(x,\nu) \bigr]\nonumber\\
&&{} +\EE_{x\sim D}
\biggl[\sum_{a\in\A}\nu(a\vert x) \EE_{r\sim
D(\,\cdot\,\vert x,a)}
\bigl[\q_k\cdot(r-\hr) \bigr] \biggr]
\nonumber
\\
&
=&\EE_{(x,a)\sim\nu} [\hr] +\EE_{(x,a,r)\sim\nu} \bigl[
\q_k\cdot(r-\hr) \bigr]
\nonumber
\\
\label{eq:exp:alt}& =&\EE_{(x,a)\sim\nu} \bigl[r^* + \bigl(\hr
-r^*\bigr) +
\q_k\cdot\bigl(r^*-\hr\bigr) \bigr]
\\
&
=&\EE_{(x,a)\sim\nu} \bigl[r^* + (1-\q_k)\Delta\bigr] .
\nonumber
\end{eqnarray}
\upqed
\end{pf}
\renewcommand{\thelemmas}{3.3}{
%leA.3 #&#
%
\begin{lemmas}
The variance of the term $\hV_{k}$ can be decomposed and bounded as follows:
\renewcommand{\theequation}{\roman{equation}}{
\setcounter{equation}{0}
%eA.2 #&#
%eA.3 #&#
%
\begin{eqnarray}
&& \hspace*{10pt}\Var^\mu_k[\hV_{k}]\\
&&\hspace*{10pt}\quad=
\VV_{x\sim D} \bigl[\EE_{a\sim\nu(\,\cdot\,
\vert x)} \bigl[r^*(x,a)\nonumber\\
&&\hspace*{10pt}\hphantom{\quad=\VV_{x\sim D} \bigl[\EE_{a\sim
\nu(\,\cdot\,
\vert x)} \bigl[}{}+ \bigl(1-
\q_k(x,a) \bigr)\Delta(x,a) \bigr] \bigr]\nonumber
\\
&&\hspace*{10pt} \qquad{} - \EE_{x\sim D} \bigl[\EE_{a\sim\nu(\,
\cdot\,\vert x)} \bigl[\q
_k(x,a)\Delta(x,a) \bigr]^2 \bigr]
\nonumber
\\
&&\hspace*{10pt}\qquad {} + \EE_{(x,a)\sim\nu} \biggl[\frac{\nu
(a\vert x)}{\hmu_k(a\vert x)} \cdot
\q_k(x,a)\cdot\VV_{r\sim D(\,\cdot\,\vert x,a)}[r] \biggr]
\nonumber
\\
&&\hspace*{10pt}\qquad {} + \EE_{(x,a)\sim\nu} \biggl[\frac{\nu
(a\vert x)}{\hmu_k(a\vert x)} \cdot
\q_k(x,a)\Delta(x,a)^2 \biggr].
\nonumber
\\
&&\hspace*{10pt} \Var^\mu_k[\hV_{k}]\\
&&\hspace*{10pt}\quad\le
\VV_{x\sim D} \bigl[r^*(x,\nu) \bigr] \nonumber\\
&&\hspace*{10pt}\qquad{}+ 2\EE_{(x,a)\sim\nu} \bigl[ \bigl\vert
\bigl(1-\q_k(x,a) \bigr)\Delta(x,a) \bigr\vert\bigr]\nonumber
\\
&&\hspace*{10pt}\qquad {} + M\EE_{(x,a)\sim\nu} \bigl[\q
_k(x,a)\nonumber\\
&&\hspace*{10pt}\hphantom{\qquad{} + M\EE_{(x,a)\sim\nu} \bigl
[}{}\cdot\EE_{r\sim D(\,\cdot\,\vert
x,a)} \bigl[ \bigl(r-\hr(x,a) \bigr)^2 \bigr] \bigr] .
\nonumber
\end{eqnarray}
}
\end{lemmas}
}
\begin{pf}
\setcounter{equation}{1}
%eA.4 #&#
%eA.5 #&#
%
\begin{eqnarray}
\E^\mu_k\bigl[\hV_{k}^2
\bigr] &=& \EE_{(x,a,r)\sim\mu_k} \biggl[ \biggl(\hat{r}(x,\nu
)+\frac{\nu
(a\vert x)}{\mu_k(a\vert x)}\cdot\q_k\nonumber\\
&&\hphantom{\EE_{(x,a,r)\sim\mu_k} \biggl[ \biggl(\hat{r}(x,\nu)+}{}
\cdot(r-\hr) \biggr)^2 \biggr]
\nonumber
\\
&=&\EE_{x\sim D} \bigl[\hat{r}(x,
\nu)^2 \bigr]\nonumber\\
&&{} +2 \EE_{(x,a,r)\sim\mu_k} \biggl[\hat{r}(x,\nu)\nonumber\\
&&\hphantom{{} +2 \EE_{(x,a,r)\sim\mu_k} \biggl[}{}\cdot
\frac{\nu
(a\vert x)}{\mu_k(a\vert x)}\cdot\q_k\cdot(r-\hr) \biggr]
\nonumber
\\
&& {} + \EE_{(x,a,r)\sim\mu_k} \biggl[\frac{\nu(a\vert x)}{\mu
_k(a\vert
x)}\nonumber\\
&&\hphantom{{} + \EE_{(x,a,r)\sim\mu_k} \biggl[}{} \cdot
\frac{\nu(a\vert x)}{\hmu_k(a\vert x)} \cdot\q_k\cdot(r-\hr)^2
\biggr]
\nonumber
\\
\label{eq:var:initial} &=& \EE_{x\sim D} \bigl[\hat{r}(x,\nu)^2
\bigr]\\
&&{} +2
\EE_{(x,a,r)\sim\nu} \bigl[\hat{r}(x,\nu)\cdot\q_k\cdot(r-\hr
) \bigr]\nonumber
\\
&& {} + \EE_{(x,a,r)\sim\nu} \biggl[\frac{\nu(a\vert x)}{\hmu
_k(a\vert x)} \cdot
\q_k\cdot(r-\hr)^2 \biggr]
\nonumber
\\
\label{eq:var:inter} &=& \EE_{(x,a)\sim\nu} \bigl[\bigl(\hat
{r}(x,\nu)-
\q_k\Delta\bigr)^2 \bigr]\\
&&{} - \EE_{(x,a)\sim\nu} \bigl[
\q_k^2\Delta^2 \bigr] + E,\nonumber
\end{eqnarray}
where $E$ denotes the term
\[
E\coloneqq\EE_{(x,a,r)\sim\nu} \biggl[\frac{\nu(a\vert x)}{\hmu
_k(a\vert x)} \cdot\q_k
\cdot(r-\hr)^2 \biggr] .
\]
To obtain an expression for the variance of $\hV_{k}$, first note that
by equation (\ref{eq:exp:alt}),
%
%eA.6 #&#
%
\begin{equation}
\label{eq:exp:initial} \E^\mu_k[\hV_{k}] =
\EE_{(x,a)\sim\nu} \bigl[\hat{r}(x,\nu) - \q_k\Delta\bigr] .
\end{equation}
Combining this with equation (\ref{eq:var:inter}), we obtain
\begin{eqnarray*}
\Var^\mu_k[\hV_{k}] &=& \VV_{(x,a)\sim\nu}
\bigl[\hat{r}(x,\nu)-\q_k\Delta\bigr]\\
&&{} - \EE_{(x,a)\sim\nu} \bigl[
\q_k^2\Delta^2 \bigr] + E
\\
&= &\VV_{x\sim D} \bigl[\EE_{a\sim\nu(\,\cdot\,\vert x)}\bigl
[\hat{r}(x,\nu)-
\q_k\Delta\bigr] \bigr]\\
&&{} + \EE_{x\sim D} \bigl[\VV_{a\sim\nu(\,\cdot\,\vert x)}
\bigl[\hat{r}(x,\nu)-\q_k\Delta\bigr] \bigr]
\\
&& {} - \EE_{x\sim D} \bigl[\VV_{a\sim\nu(\,\cdot\,\vert x)}[\q
_k\Delta]
\bigr] \\
&&{}- \EE_{x\sim D} \bigl[\EE_{a\sim\nu(\,\cdot\,\vert x)}[\q
_k\Delta
]^2 \bigr] + E
\\
&=& \VV_{x\sim D} \bigl[\EE_{a\sim\nu(\,\cdot\,\vert x)}\bigl
[r^*+(1-\q_k)
\Delta\bigr] \bigr] \\
&&{}+ \EE_{x\sim D} \bigl[\VV_{a\sim\nu(\,\cdot\,\vert x)}[
\q_k\Delta] \bigr]
\\
&& {} - \EE_{x\sim D} \bigl[\VV_{a\sim\nu(\,\cdot\,\vert x)}[\q
_k\Delta]
\bigr]\\
&&{} - \EE_{x\sim D} \bigl[\EE_{a\sim\nu(\,\cdot\,\vert x)}[\q
_k\Delta
]^2 \bigr] + E
\\
&=& \VV_{x\sim D} \bigl[\EE_{a\sim\nu(\,\cdot\,\vert x)}\bigl
[r^*+(1-\q_k)
\Delta\bigr] \bigr]\\
&&{} - \EE_{x\sim D} \bigl[\EE_{a\sim\nu(\,\cdot\,\vert x)}[
\q_k\Delta]^2 \bigr] + E .
\end{eqnarray*}
We now obtain part (i) of the lemma
by decomposing the term $E$:
\begin{eqnarray*}
E&=& \EE_{(x,a,r)\sim\nu} \biggl[\frac{\nu(a\vert x)}{\hmu
_k(a\vert x)} \cdot\q_k\cdot
\bigl(r-r^*\bigr)^2 \biggr]\\
&&{} + \EE_{(x,a)\sim\nu} \biggl[
\frac{\nu(a\vert x)}{\hmu_k(a\vert x)} \cdot\q_k\cdot\bigl
(r^*-\hr\bigr)^2
\biggr]
\\
&=& \EE_{(x,a)\sim\nu} \biggl[\frac{\nu(a\vert x)}{\hmu_k(a\vert
x)} \cdot\q_k\cdot
\VV_{r\sim D(\,\cdot\,\vert x,a)}[r] \biggr] \\
&&{}+ \EE_{(x,a)\sim\nu} \biggl[\frac{\nu(a\vert x)}{\hmu
_k(a\vert x)} \cdot
\q_k\Delta^2 \biggr] .
\end{eqnarray*}
To prove part (ii) of the lemma, first note that
\begin{eqnarray*}
\hat{r}(x,\nu)^2 &=& \bigl(r^*(x,\nu)+\EE_{a\sim\nu(\,\cdot\,
\vert x)} \bigl[\Delta
(x,a) \bigr] \bigr)^2
\\
&=&r^*(x,\nu)^2+2r^*(x,\nu)\EE_{a\sim\nu(\,\cdot\,\vert x)} \bigl
[\Delta(x,a)
\bigr]\\
&&{} +\EE_{a\sim\nu(\,\cdot\,\vert x)} \bigl[\Delta(x,a) \bigr]^2
\\
&=&r^*(x,\nu)^2+2\hat{r}(x,\nu)\EE_{a\sim\nu(\,\cdot\,\vert x)}
\bigl[\Delta(x,a)
\bigr]\\
&&{} -\EE_{a\sim\nu(\,\cdot\,\vert x)} \bigl[\Delta(x,a) \bigr]^2
\\
&\le& r^*(x,\nu)^2+2\hat{r}(x,\nu)\EE_{a\sim\nu(\,\cdot\,\vert
x)} \bigl[
\Delta(x,a) \bigr] .
\end{eqnarray*}
Plugging this in equation (\ref{eq:var:initial}), we obtain
%
%eA.7 #&#
%
\begin{eqnarray}
\E^\mu_k\bigl[\hV_{k}^2
\bigr] & =& \EE_{x\sim D} \bigl[\hat{r}(x,\nu)^2 \bigr]\nonumber
\\
&&{}+ 2
\EE_{(x,a,r)\sim\nu} \bigl[\hat{r}(x,\nu)\cdot\q_k\cdot(r-\hr
) \bigr]
+ E
\nonumber
\\
&\le& \EE_{x\sim D} \bigl[r^*(x,\nu)^2 \bigr] \nonumber\\
&&{}+ 2
\EE_{x\sim D} \bigl[\hat{r}(x,\nu)\EE_{a\sim\nu(\,\cdot\,\vert
x)}[\Delta] \bigr]
\nonumber
\\
&&{} + 2 \EE_{(x,a)\sim\nu} \bigl[\hat{r}(x,\nu)\cdot(-
\q_k)\cdot\Delta\bigr] + E
\nonumber
\\
\label{eq:bound:1} &=& \EE_{x\sim D} \bigl[r^*(x,\nu)^2 \bigr]\\
&&{}+ 2
\EE_{(x,a)\sim\nu} \bigl[\hat{r}(x,\nu)\cdot(1-\q_k)\cdot
\Delta
\bigr] + E .\nonumber
\end{eqnarray}
On the other hand, equation (\ref{eq:exp:initial}) can be rewritten as
\[
\E^\mu_k[\hV_{k}] = \EE_{(x,a)\sim\nu}
\bigl[r^*(x,\nu) +(1- \q_k)\Delta\bigr] .
\]
Combining with equation (\ref{eq:bound:1}), we obtain
\begin{eqnarray*}
\Var^\mu_k[\hV_{k}] &\le& \VV_{x\sim D}
\bigl[r^*(x,\nu) \bigr] \\
&&{}+ 2\EE_{(x,a)\sim\nu} \bigl[\hat{r}(x,\nu)\cdot(1-
\q_k)\Delta\bigr]
\\
&& {} - 2\EE_{x\sim D} \bigl[r^*(x,\nu) \bigr] \EE_{(x,a)\sim\nu
} \bigl[(1-
\q_k)\Delta\bigr] \\
&&{}- \EE_{(x,a)\sim\nu} \bigl[(1- \q_k)
\Delta\bigr]^2 + E
\\
&\le& \VV_{x\sim D} \bigl[r^*(x,\nu) \bigr]\\
&&{} + 2\EE_{(x,a)\sim\nu} \bigl[
\bigl(\hat{r}(x,\nu)-\tfrac{1}{2} \bigr) (1-\q_k)\Delta
\bigr]
\\
&& {} - 2\EE_{x\sim D} \bigl[r^*(x,\nu)-\tfrac{1}{2} \bigr]
\EE_{(x,a)\sim
\nu} \bigl[(1-\q_k)\Delta\bigr]\\
&&{} + E
\\
&\le& \VV_{x\sim D} \bigl[r^*(x,\nu) \bigr] + \EE_{(x,a)\sim\nu
} \bigl[
\bigl\vert(1-\q_k)\Delta\bigr\vert\bigr]\\
&&{} + \bigl\llvert\EE_{(x,a)\sim\nu}
\bigl[(1-\q_k)\Delta\bigr]\bigr\rrvert+ E,
\end{eqnarray*}
where the last inequality follows by H\"older's inequality
and the observations that $\vert\hr-1/2\vert\le1/2$ and $\vert
r^*-1/2\vert\le1/2$.
Part (ii) now follows by the bound
\begin{eqnarray*}
E &=& \EE_{(x,a,r)\sim\nu} \biggl[\frac{\nu(a\vert x)}{\hmu
_k(a\vert
x)} \cdot\q_k\cdot(r-
\hr)^2 \biggr] \\
&\le& M\EE_{(x,a)\sim\nu} \bigl[\q_k
\EE_{r\sim D(\,\cdot\,\vert
x,a)} \bigl[(r-\hr)^2 \bigr] \bigr].
\end{eqnarray*}
\upqed
\end{pf}

%sA #&#
\section{Freedman's Inequality}
\label{app:freedman}

The following is a corollary of Theorem~1 of
\citet{EXP4P}. It can be viewed as a version of Freedman's
inequality \citeauthor{Martingale}'s (\citeyear{Martingale}).
Let $y_{1},\ldots,y_{n}$ be a sequence of real-valued random variables.
Let $\E_k$ denote
$\E[{}\cdot{}| y_1,\ldots,y_{k-1}]$
and $\Var_k$ conditional variance.
%
%thA.1 #&#
%
\begin{theorem}
\label{thm:freedman}
Let $V,D\in\mathbb{R}$ such that
\[
\sum_{k=1}^n \Var_k[y_k]\le V,
\]
and for all $k$,
$\llvert y_k-\E_k[y_k]\rrvert\le D$. Then for any $\delta>0$,
with probability at least $1-\delta$,
\[
\Biggl\llvert\sum_{k=1}^n
y_k -\sum_{k=1}^n
\E_k[y_k]\Biggr\rrvert\le2\max\bigl\{D\ln(2/\delta),
\sqrt{V\ln(2/\delta)} \bigr\} .
\]
\end{theorem}

%sA #&#
\section{Improved Finite-sample Error Bound}
\label{app:error}

In this appendix, we analyze the error of $\hV_\DR$ in estimating
the value of a stationary policy $\nu$. We generalize the analysis of
Section~\ref{sec:error}
by replacing conditions on the ranges of variables by conditions
on the moments.

For a function $f\dvtx\X\times\A\to\R$ and $1\le p<\infty$, we define
the $L_p(\nu)$ norm as usual:
\[
\Vert f\Vert_{p,\nu} = \E_{(x,a)\sim\nu} \bigl[\bigl\vert f(x,a)
\bigr\vert^p \bigr]^{1/p} .
\]
For $p=\infty$, $\Vert f\Vert_{\infty,\nu}$ is the essential
supremum of $\vert f\vert$ under $\nu$.

As in Section~\ref{sec:error}, we first simplify Lemmas \ref
{LEMMA:RANGE}--\ref{LEMMA:VAR}, and then
apply Freedman's inequality to obtain a specific error bound.
%
%leA.1 #&#
%
\begin{lemma}
\label{lemma:bound:stationary:holder}
Let $1\le p,q\le\infty$ be such that $1/p+1/q=1$. Assume there
are finite constants $M, e_{\hr},\delta_\Delta,\delta_\q,\qmax\ge
0$ such
that with probability one under $\mu$, for all $k$:
\begin{eqnarray*}
\nu(a_k\vert x_k)/\hmu_k(a_k
\vert x_k)&\le& M,\\
\Vert\Delta\Vert_{q,\nu}&\le&\delta_\Delta,
\\
\Vert1-\q_k\Vert_{p,\nu}&\le&\delta_\q,
\\
\Vert\q_k\Vert_{p,\nu}&\le&\qmax,
\\
\EE_{(x,a)\sim\nu} \bigl[\EE_{r\sim D(\,\cdot\,\vert x,a)} \bigl
[ \bigl(\hr(x,a)-r
\bigr)^2 \bigr]^q \bigr]^{1/q}&\le&
e_{\hr} .
\end{eqnarray*}
Then with probability one under $\mu$, for all $k$:
\begin{eqnarray*}
\bigl\llvert\E^\mu_k[\hV_{k}]-V\bigr
\rrvert&\le&\delta_\q\delta_\Delta,
\\
\Var^\mu_k[\hV_{k}] &\le& \Var_{x\sim D}
\bigl[r^*(x,\nu)\bigr] +2\delta_\q\delta_\Delta+ M\qmax
e_{\hr} .
\end{eqnarray*}
\end{lemma}
\begin{pf}
%The range bound is just Lemma~\ref{lemma:range}.
The bias and variance
bound follow from Lemma~\ref{lemma:exp}
and Lemma~\ref{LEMMA:VAR}(ii), respectively, by H\"ol\-der's inequality.
\end{pf}
%
%thA.2 #&#
%
\begin{theorem}
\label{thm:error:holder}
If assumptions of Lemma~\textup{\ref{lemma:bound:stationary:holder}}
hold, then
with probability at least $1-\delta$,
{\fontsize{10.6}{12.6}\selectfont{
\begin{eqnarray*}
&&\hspace*{-4pt}\vert\hV_\DR- V \vert\\
&&\hspace*{-5pt}\quad\le\delta_\q\delta_\Delta\\
&&\hspace*{-6pt}\qquad{}+ 2\max\biggl\{ \frac{(1+M)\ln(2/\delta)}{n},
\\
&&\hspace*{-6pt}\qquad\sqrt{\frac{ (\Var_{x\sim D}[r^*(x,\nu)]
+2\delta_\q\delta
_\Delta+ M\qmax e_{\hr} )\ln(2/\delta)}{n}} \biggr\} .
\end{eqnarray*}
}}
\end{theorem}
\begin{pf}
The proof follows by Freedman's inequality (Theorem~\ref{thm:freedman}
in Appendix~\ref{app:freedman}), applied to random variables $\hV
_{k}$, whose range and variance are bounded using Lemma~\ref
{LEMMA:RANGE} and \ref{lemma:bound:stationary:holder}.
\end{pf}

%sA #&#
\section{Direct Loss Minimization}
\label{app:dlm}

Given cost-sensitive multiclass classification data $\{(x,l_1,\ldots
,l_K)\}$,
%an algorithm by \namecite{McAllester11Direct} may be instantiated
we perform approximate gradient descent on
the policy loss (or classification error). In the experiments of
Section~\ref{sec:class}, policy $\nu$ is specified by $K$ weight
vectors $\theta_1,\ldots,\theta_K$. Given $x\in\X$, the policy
predicts as follows: $\nu(x)=\argmax_{a\in\{1,\ldots,K\}}\{x\cdot
\theta_{a}\}$.

To optimize $\theta_a$, we adapt the ``toward-better'' version of the
direct loss minimization method of \citet{McAllester11Direct} as
follows: given any data point $(x,l_1,\ldots,l_K)$ and the current
weights $\theta_a$, the weights are adjusted by
\begin{eqnarray*}
\theta_{a_1} &\leftarrow&\theta_{a_1} + \eta x,\\
\theta_{a_2} &\leftarrow&\theta_{a_2} - \eta x,
\end{eqnarray*}
where $a_1 = \argmax_{a} \{x\cdot\theta_{a}-\epsilon
l_{a} \}$, $a_2 = \argmax_{a} \{x\cdot\theta_{a} \}
$, $\eta\in(0,1)$ is a decaying learning rate, and $\epsilon>0$ is
an input parameter.

% MD: batch updates = mini-batch or is this essentially subgradient
%method?
%
For computational reasons, we actually perform batch updates rather than
incremental updates. Updates continue until the weights converge. We
found that the
learning rate $\eta=t^{-0.3}/2$, where $t$ is the batch iteration,
worked well across all datasets. The parameter $\epsilon$ was fixed to
$0.1$ for all datasets.

Furthermore, since the policy loss is not convex in the weight vectors,
we repeat the algorithm $20$ times with randomly perturbed starting
weights and then return the best run's weight according to the learned
policy's loss in the training data. We also tried using a holdout
validation set for choosing the best weights out of the $20$
candidates, but did not observe benefits from doing so.

%sA #&#
\section{Filter Tree}
\label{app:tree}

The Filter Tree (Beygelzimer, Langford and Ravi\-kumar, \citeyear
{Beygelzimer08Multiclass}) is a reduction from
multiclass cost-sensitive classification to binary classification. Its
input is of the same form as for Direct Loss Minimization, but its
output is a Filter Tree: a decision tree, where each inner node
is itself implemented by some binary classifier (called base
classifier), and leaves correspond to
classes of the original multiclass problem. As base classifiers
we used J48 decision trees
implemented in Weka 3.6.4 (\citeauthor{Hall09Weka}, \citeyear
{Hall09Weka}). Thus,
there are 2-class decision trees in the nodes, with the nodes arranged
as per a Filter Tree. Training in a Filter Tree proceeds bottom-up,
but the classification in a trained Filter Tree proceeds root-to-leaf,
with the running time logarithmic in the number of classes. We did not test
the all-pairs Filter Tree, which classifies examples
in the time linear in the number of classes, similar to DLM.

%sA #&#
\section{\texorpdfstring{Proofs of Lemmas \protect\ref{LEMMA:COND} and \protect\ref{LEMMA:MARGINAL}}{Proofs of Lemmas 5.2 and 5.3}}
\label{app:proofs2}
\renewcommand{\thelemmas}{5.2}{
%leA.1 #&#
%
\begin{lemmas}
Let $t\le T$, $k\ge1$ and let $z_{k-1}$ be such that the $k$th
exploration sample marks the beginning of the $t$th block,
that is, $\kappa(t-1)=k-1$.
Let $h_{t-1}$ and $c_{t}$ be the target history and
acceptance rate multiplier induced by $z_{k-1}$.
Then:
%
%eA.1 #&#
%eA.2 #&#
%
\begin{eqnarray}
\label{eq:lemma:prob}
&& \sum_{x,a}\bigl\llvert
\Pr^\mu_k[x_{\kappa(t)}=x,a_{\kappa(t)}=a] -
\pi_t(x,a) \bigr\rrvert\\
&&\quad \le\frac{2\eps}{1-\eps} ,\nonumber
\\
\label{eq:lemma:exp}
&&\bigl\llvert c_t\E^\mu_k[
\hV_{B(t)}] - \E^\pi_t[r]\bigr\rrvert\le
\frac{\eps}{1-\eps} .
\end{eqnarray}
\end{lemmas}
}
\begin{pf}
We begin by showing
equation (\ref{eq:lemma:prob}). Consider the $m$th exploration
sample $(x,a)\sim\mu_m$ and assume that this sample
is in the $t$th block. The probability
of accepting this sample is
\begin{eqnarray*}
&&\PP_{u\sim\mu_m(\,\cdot\,\vert x,a)} \biggl[u \le\frac{c_t\pi
_t(a\vert x)}{\mu_m(a\vert x)} \biggr] \\
&&\quad= \ind
\bigl[(x,a)\in\event_m \bigr] + \frac{c_t\pi_t(a\vert x)}{\mu
_m(a\vert x)} \ind\bigl[(x,a)
\notin\event_m \bigr] ,
\end{eqnarray*}
where $\ind[\,\cdot\,]$ is the indicator function equal to $1$ when
its argument is true and $0$ otherwise.
The probability of seeing and accepting a sample $(x,a)$ from
$\mu_m$ is
\begin{eqnarray*}
&&\accept_m(x,a) \\
&& \quad\coloneqq\mu_m(x,a) \biggl(\ind
\bigl[(x,a)\in\event_m \bigr] \\
&&\hphantom{\quad\coloneqq\mu_m(x,a) \biggl(}{}+ \frac
{c_{t}\pi_t(a\vert x)}{\mu_m(a\vert x)} \ind\bigl[(x,a)
\notin\event_m \bigr] \biggr)
\\
&&\quad = \mu_m(x,a)\ind\bigl[(x,a)\in\event_m \bigr]\\
&&\qquad{} +
c_{t}\pi_t(x,a)\ind\bigl[(x,a)\notin\event_m
\bigr]
\\
&&\quad = c_{t}\pi_t(x,a) \\
&&\qquad{}- \bigl(c_{t}
\pi_t(x,a) - \mu_m(x,a) \bigr)\ind\bigl[(x,a)\in
\event_m \bigr]
\end{eqnarray*}
and the marginal probability of accepting a sample from $\mu_m$ is
\begin{eqnarray*}
\accept_m(*) &\coloneqq& \sum_{x,a}
\accept_m(x,a)\\
& =& c_{t} - c_{t}\eps_m
= c_{t}(1-\eps_m) .
\end{eqnarray*}
In order to accept the $m$th exploration sample, samples $k$ through
$m-1$ must
be rejected. The probability of eventually accepting $(x,a)$,
conditioned on
$z_{k-1}$ is therefore
%
%eA.3 #&#
%eA.4 #&#
%
\begin{eqnarray}
&& \Pr^\mu_k (x_{\kappa(t)}=x,a_{\kappa(t)}=a
)
\nonumber
\\
&&\quad= \E^\mu_k \Biggl[\sum
_{m\ge k}^\infty\accept_m(x,a)\prod
_{k'=k}^{m-1}\bigl(1-\accept_{k'}(*)\bigr)
\Biggr]\hspace*{-17pt}
\nonumber
\\
\label{eq:Prk:1} &&\quad= c_{t}\pi_t(x,a)\\
&&\qquad{}\cdot\E^\mu_k\Biggl[\sum_{m\ge k}^\infty\prod
_{k'=k}^{m-1}\bigl(1-\accept_{k'}(*)\bigr)
\Biggr]\nonumber
\\
\label{eq:Prk:2}&&\qquad {} -\E^\mu_k \Biggl[ \sum
_{m\ge k}^\infty\bigl(c_{t}
\pi_t(x,a) - \mu_m(x,a) \bigr) \\
&&\hphantom{\qquad {} -\E^\mu_k \Biggl[\sum
_{m\ge k}^\infty}{}\cdot\ind\bigl[(x,a)\in
\event_m \bigr] \nonumber\\
&& \hspace*{76pt}{}\cdot
\prod_{k'=k}^{m-1}\bigl(1-\accept_{k'}(*)
\bigr) \Biggr] .\nonumber
\end{eqnarray}
To bound $\llvert\Pr^\mu_k[x_{\kappa(t)}=x,a_{\kappa
(t)}=a]-\pi_t(x,a)\rrvert$
and prove equation (\ref{eq:lemma:prob}), we first need to bound
equations (\ref{eq:Prk:1}) and (\ref{eq:Prk:2}).
Note that from the definition of $\event_m$,
the expression inside the expectation of equation (\ref{eq:Prk:2})
is always
nonnegative. Let $E_1(x,a)$
denote the expression in equation (\ref{eq:Prk:1}) and $E_2(x,a)$ the
expression in
equation (\ref{eq:Prk:2}). We bound $E_1(x,a)$ and $E_2(x,a)$
separately, using bounds
$0\le\eps_m\le\eps$:\vspace*{2pt}
\begin{eqnarray*}
E_1(x,a) & =& c_{t}\pi_t(x,a)
\E^\mu_k \Biggl[\sum_{m\ge k}^\infty
\prod_{k'=k}^{m-1}\bigl(1-\accept_{k'}(*)
\bigr) \Biggr]
\\
& \le&c_{t}\pi_t(x,a)\E^\mu_k
\Biggl[\sum_{m\ge k}^\infty\prod
_{k'=k}^{m-1}\bigl(1-c_{t}(1-\eps)\bigr)
\Biggr]
\\
& =& \frac{\pi_t(x,a)}{1-\eps},
\\
E_1(x,a) & \ge& c_{t}\pi_t(x,a)
\E^\mu_k \Biggl[\sum_{m\ge k}^\infty
\prod_{k'=k}^{m-1}(1-c_{t}) \Biggr]
\\
& =&\pi_t(x,a),
\\
E_2(x,a) & =& \E^\mu_k \Biggl[ \sum
_{m\ge k}^\infty\bigl(c_{t}
\pi_t(x,a) - \mu_m(x,a) \bigr)\\
&&\hphantom{\E^\mu_k \Biggl[}{}\cdot\ind\bigl[(x,a)\in
\event_m \bigr] \\
&&\hphantom{\E^\mu_k \Biggl[}{}\cdot\prod_{k'=k}^{m-1}
\bigl(1-\accept_{k'}(*)\bigr) \Biggr]
\\
& \le& \E^\mu_k \Biggl[ \sum
_{m\ge k}^\infty\bigl(c_{t}
\pi_t(x,a) - \mu_m(x,a) \bigr)\\
&&\hphantom{\E^\mu_k \Biggl[}{}\cdot\ind\bigl[(x,a)\in
\event_m \bigr] \\
&&\hspace*{50pt}{}\cdot\bigl(1-c_{t}(1-\eps)
\bigr)^{m-k} \Biggr].
\end{eqnarray*}
Now we are ready to prove equation (\ref{eq:lemma:prob}):
\begin{eqnarray*}
&& \sum_{x,a} \vert\Pr^\mu_k[x_{\kappa(t)}=x,a_{\kappa(t)}=a]
- \pi_t(x,a) \vert
\\
&&\quad = \sum_{x,a} \vert E_1(x,a)-
\pi_t(x,a)-E_2(x,a) \vert
\\
& &\quad \le\sum_{x,a} \vert E_1(x,a)-
\pi_t(x,a) \vert+\sum_{x,a}
E_2(x,a)
\\
&&\quad \le\sum_{x,a}\frac{\pi_t(x,a)\eps}{1-\eps}
\\
& &\qquad{} +\E^\mu_k \Biggl[ \sum
_{m\ge k}^\infty\sum_{x,a}
\bigl(c_{t}\pi_t(x,a) - \mu_m(x,a) \bigr)
\\
&&\hphantom{\qquad{} +\E^\mu_k \Biggl[} {}
\cdot\ind\bigl[(x,a)\in\event_m \bigr]
\bigl(1-c_{t}(1-\eps)\bigr)^{m-k} \Biggr]
\\
& &\quad = \frac{\eps}{1-\eps} +\E^\mu_k \Biggl[\sum
_{m\ge k}^\infty c_{t}\eps_m
\bigl(1-c_{t}(1-\eps)\bigr)^{m-k} \Biggr]
\\
& &\quad \le\frac{2\eps}{1-\eps}
\end{eqnarray*}
proving equation (\ref{eq:lemma:prob}).

Let $\reach_m$ denote the indicator of the event that the $m$th sample
is in block $t$ (i.e., samples
$k,k+1,\ldots,m-1$ are rejected).
Then
%
%eA.5 #&#
%
\begin{eqnarray}
\E^\mu_k [\hV_{B(t)} ] &=&\sum
_{m=k}^\infty\E^\mu_k [
\hV_{m}\reach_m ]
\nonumber
\\
&
=&\sum_{m=k}^\infty
\E^\mu_k \bigl[\E^\mu_m [
\hV_{m}\reach_m ] \bigr]
\nonumber
\\
\label{eq:Et:Rt+1} &=&\sum_{m=k}^\infty
\E^\mu_k \bigl[\reach_m\E^\mu_m[
\hV_{m}] \bigr] ,
\end{eqnarray}
where equation (\ref{eq:Et:Rt+1}) follows because the event of
reaching the
$m$th sample depends only on the
preceding samples, and hence it is a deterministic function of
$z_{m-1}$. Plugging Lemma~\ref{lemma:exp} in equation (\ref
{eq:Et:Rt+1}), we obtain
\begin{eqnarray*}
&&c_{t}\E^\mu_k [\hV_{B(t)} ]\\
&&\quad
=c_{t}\EE_{r\sim\pi_t}[r] \sum_{m=k}^\infty
\E^\mu_k[\reach_m]
\\
&&\quad =c_{t}\EE_{r\sim\pi_t}[r] \E^\mu_k
\Biggl[\sum_{m=k}^\infty\prod
_{k'=k}^{m-1}\bigl(1-\accept_{k'}(*)\bigr)
\Biggr]
\end{eqnarray*}
(because $\EE_{r\sim\pi_t}[r]$ is a deterministic function of $z_{k-1}$).
This can be bounded, similarly as before, as
\[
\EE_{r\sim\pi_t}[r] \le c_{t}\E^\mu_k [
\hV_{B(t)} ] \le\frac{\EE_{r\sim\pi_t}[r]}{1-\eps}
\]
yielding equation (\ref{eq:lemma:exp}).
\end{pf}
\renewcommand{\thelemmas}{5.3}
%leA.2 #&#
%
\begin{lemmas}
\[
\sum_{h_T} \bigl\vert\hpi(h_T) - \pi(h_T)\bigr\vert
\le
(2\eps T) / (1-\eps)
.
\]
\end{lemmas}
\begin{pf}
We prove the lemma by induction and the triangle inequality
(essentially following Kakade, Kearns and Langford, \citeyear{KKL03}). The
lemma holds for $T=0$
since there is only one empty
history (and hence both $\hpi$ and $\pi$ are point distributions
over $h_0$). Now assume the lemma holds for $T-1$. We
prove it for $T$:
\begin{eqnarray*}
&& \sum_{h_T}\bigl\vert\hpi(h_T) -
\pi(h_T) \bigr\vert
\nonumber
\\
&&\quad
=\sum_{h_{T-1}} \sum
_{(x_T,a_T,r_T)} \bigl\vert\hpi(h_{T-1})\hpi_{T}(x_T,a_T,r_T)\\
&&\hphantom{\quad
=\sum_{h_{T-1}} \sum
_{(x_T,a_T,r_T)} \vert}{}- \pi(h_{T-1})\pi_{T}(x_T,a_T,r_T)
\bigr\vert
\nonumber
\\
& &\quad
\le\sum_{h_{T-1}} \sum
_{(x_T,a_T,r_T)} \bigl( \bigl\vert\hpi(h_{T-1})\hpi
_{T}(x_T,a_T,r_T)\\
&&\hphantom{\quad
\le\sum_{h_{T-1}} \sum
_{(x_T,a_T,r_T)} \bigl( \vert}{}-\hpi(h_{T-1})\pi_{T}(x_T,a_T,r_T)
\bigr\vert
\nonumber
\\
&&\hphantom{\quad
\le\sum_{h_{T-1}} \sum
_{(x_T,a_T,r_T)} \bigl(} {}
+ \bigl\vert\hpi(h_{T-1})\pi_T(x_T,a_T,r_T)\\
&&\hphantom{\quad
\le\sum_{h_{T-1}} \sum
_{(x_T,a_T,r_T)} \bigl( \vert{}+}{}- \pi(h_{T-1})\pi_T(x_T,a_T,r_T)
\bigr\vert\bigr)
\nonumber
\\
& &\quad
= \EE_{h_{T-1}\sim\hpi} \biggl[\sum_{(x_T,a_T,r_T)}
\bigl\vert\hpi_T(x_T,a_T,r_T)\\
&&\hphantom{\quad
= \EE_{h_{T-1}\sim\hpi} \biggl[\sum_{(x_T,a_T,r_T)}
}{} -
\pi_T(x_T,a_T,r_T) \bigr\vert\biggr]
\nonumber
\\
&&\qquad {}
+\sum_{h_{T-1}} \bigl\vert\hpi(h_{T-1})
- \pi(h_{T-1}) \bigr\vert
\nonumber
\\
& &
\quad\le\frac{2\eps}{1-\eps} + \frac{2\eps(T-1)}{1-\eps} =
\frac{2\eps T}{1-\eps} .
\nonumber
\end{eqnarray*}
\upqed
\end{pf}

%sA #&#
\section{Progressive Validation Policy}
\label{app:PE}

In Section~\ref{sec:class-opt}, we showed how the stationary DR
estimator can be
used not only for policy evaluation, but also for policy optimization
by transforming the contextual bandit problem into
a cost sensitive classification problem.

In this appendix, we show how the \emph{nonstationary} DR estimator,
when applied to an online learning algorithm, can also be used to
obtain a high-performing \emph{stationary} policy. The value of this
policy concentrates around the average per-step reward estimated for
the online learning algorithm. Thus, to the extent that the online
algorithm achieves a high reward, so does this stationary policy.
The policy is constructed using the ideas behind
the ``progressive validation'' error bound (\citeauthor{BKL99},
\citeyear{BKL99}),
and hence we call it a ``progressive validation policy.''

Assume that the algorithm DR-ns successfully terminates
after generating $T$ blocks. The
\emph{progressive validation policy}
is the randomized stationary policy $\piPE$ defined as
\[
\piPE(a\vert x) \coloneqq\sum_{t=1}^{T}
\frac{c_t\vert B(t)\vert}{C}\pi(a\vert x,h_{t-1}) .
\]
Conceptually, this policy first picks among the histories $h_0,\ldots,h_{T-1}$
with probabilities $c_1\vert B(1)\vert/C, \ldots,\break
c_t\vert
B(T)\vert/C$, and then executes the
policy $\pi$ given the chosen history.
We extend $\piPE$ to a distribution over triples
\[
\piPE(x,a,r) = D(x)\piPE(a\vert x) D(r\vert x,a) .
\]

We will show that the average reward estimator
$\hat{V}^{\mathrm{avg}}_\DRns$ returned by our algorithm estimates the
expected reward of $\piPE$ with an error
$O(1/\sqrt{N})$ where $N$ is the number of exploration
samples used to generate $T$ blocks. Thus, assuming
that the nonstationary policy $\pi$ improves with more data,
we expect to obtain
the \emph{best-performing} progressive validation policy with the
\emph{most accurate}
value estimate by running the algorithm DR-ns on
all of the exploration data.

The error bound in the theorem below
is proved by analyzing range and variance of $\hV_{k}$ using
Lemma~\ref{lemma:bound:stationary}.
The theorem relies on the following conditions
(mirroring the assumptions of Lemma~\ref{lemma:bound:stationary}):
\begin{itemize}
\item There is a constant $M>0$ such that $\pi_t(a_k\vert x_k)
/p_k\le M$.
\item There is a constant $e_{\hr}>0$ such that\newline $\EE_{(x,a)\sim\pi
_t} [\E_D [(\hr-r)^2\bigm\vert x,a ] ]\le
e_{\hr}$.
\item There is a constant $v_r>0$ such that\newline $\VV_{x\sim D} [\EE
_{r,a\sim\pi_t(\,\cdot,\cdot\,\vert x)}[r] ]\le v_r$.
\end{itemize}
These conditions ensure boundedness of density ratios,
squared prediction error of rewards,
and variance of a conditional expected reward, respectively.
It should be noted that, since rewards are assumed to be in $[0,1]$,
one can always choose $e_{\hr}$ and $v_r$ that are no greater than $1$.
%
%thA.1 #&#
%
\begin{theorem}
Let $N$ be the number of exploration samples used to generate $T$
blocks, that is,
$N=\sum_{t=1}^T\vert B(t)\vert$.
%Assume there are finite constants $M, e_{\hr},\varbound\ge0$ such that
Assume the above conditions hold for all $k$ and $t$ (and all histories
$z_{k-1}$ and $h_{t-1}$).
%
% \frac{\pi_k(a_k\given x_k)}{p_k}\le M
%,
%
% \EE_{(x,a)\sim\pi_k}\Bracks{\E_D\Bracks{(\hr-r)^2\bigGiven x,a}}\le
%e_{\hr}
%,
%
% \VV_{x\sim D}\Bracks{\EE_{r,a\sim\pi_k(\cdot,\cdot\given x)}[r]}
% \le\varbound
%.
%
Then, with probability at least $1-\delta$,\vspace*{-1pt}
\begin{eqnarray*}
&&\bigl\llvert\hat{V}^{\mathrm{avg}}_\DRns- \EE_{r\sim\piPE
}[r]\bigr
\rrvert\\
&&\quad \le\frac{Nc_{\max}}{C} \\
&&\qquad{}\cdot2\max\biggl\{ \frac{(1+M)\ln(2/\delta)}{N}, \\
&&\hphantom{\qquad{}\cdot2\max\biggl\{}\sqrt{
\frac{ (v_r + M e_{\hr} )\ln(2/\delta)}{N}} \biggr\} .
\end{eqnarray*}
\end{theorem}
\begin{pf}
The proof follows by Freedman's inequality (Theorem~\ref{thm:freedman}
in Appendix~\ref{app:freedman}),
applied to random variables $c_t \hV_{k}$, whose range and variance
can be bounded using
Lemma~\ref{lemma:bound:stationary} and the bound $c_t\le c_{\max}$.
In applying
Lemma~\ref{lemma:bound:stationary}, note
that $\delta_\q=0$ and $\qmax=1$, because $\hmu_k=\mu_k$.
\end{pf}
\end{appendix}

\section*{Acknowledgments}
We thank the editors and reviewers for their vast amount of patience
and efforts that have improved the paper substantially.
%suskaldyti doi

% imsref loaded by jurgita.kaciuliene, 2014-10-21 08:57:03
%

%
\end{document}